\documentclass[journal]{IEEEtran}
\usepackage{amsfonts}
\usepackage{amssymb}
\usepackage{eurosym}
\usepackage{cite}
\usepackage{graphicx}
\usepackage{epstopdf}
\usepackage{amsmath}
\usepackage{tikz,lipsum}
\usepackage{caption}
\usepackage[T1]{fontenc}
\usepackage{amsthm}
\usepackage{mathrsfs}
\usepackage{color}
\usepackage{stfloats}
\usepackage{xcolor, etoolbox}
\usepackage{subcaption}
\usepackage{comment}
\usepackage{algcompatible}
\usepackage[ruled]{algorithm2e}
\usepackage{pifont}  % in the preamble
\usepackage{multirow}
\IEEEoverridecommandlockouts

\newcommand{\xmark}{\ding{55}}%

\begin{document}

\title{Pinching Antenna Systems (PASS): Enabling Reconfigurable and Controllable Wireless Channels -- A Comprehensive Survey}
\author{Elmehdi Illi, \IEEEmembership{Member, IEEE}, and Marwa Qaraqe,
\IEEEmembership{Senior
Member, IEEE}\thanks{%
E. Illi and M. Qaraqe are with the College of Science and
Engineering, Hamad Bin Khalifa University, Doha, Qatar. (e-mails:
elmehdi.illi@ieee.org, mqaraqe@hbku.edu.qa.)} }
\maketitle

\begin{abstract}
The evolution of wireless networks constantly brings new paradigms into action for consideration in upcoming generations. To this end, the sixth generation of wireless networks (6G) anticipates the development of several data rate-hungry applications, in addition to a forecast growth of sensing-centric applications. Such an evolution, however, is unbalanced on the other side by the accentuated scarcity of spectrum, which opens up urgent needs to develop spectrum-efficient communication and sensing techniques. Due to the inability of the traditional multi-antenna schemes to enhance a wireless channel quality, an increasing interest has been paid to wireless channel-altering schemes, such as reconfigurable intelligent surfaces and movable antennas. Recently, a new technique in this category, called pinching antennas (PAs), was introduced and tested. PA systems (PASS) are based on extending the reach of a base station by connecting its radio-frequency chains to long waveguides, on which one or many radiating antennas are pinched at custom positions of interest. Thus, such a technique can provide a means of overcoming several unfavorable channel conditions, such as the absence of a line-of-sight and increased free-space path loss. Importantly, such a channel-tuning feature can provide notable enhancements in terms of sensing, network coverage, data rate, and resilience against eavesdropping. In this work, we provide a comprehensive review of the research work performed on PASS, designed on the basis of various system design objectives, such as the network coverage or data rate, information-theoretic secure transmission, sensing, integrated sensing and communication, and energy efficiency. A categorization of the surveyed research work is established with a comparison between the various PASS schemes presented. Several takeaways are illustrated on the proposed schemes' potential and limitations, along with several challenges and directions forward discussed, in terms of future deployment and implementation.  
\end{abstract}

\begin{IEEEkeywords}
6G, network coverage, data rate, eavesdropping, integrated sensing and communication (ISAC), pinching antennas, and spectrum efficiency.
\end{IEEEkeywords}

\section{Introduction}
\subsection{Background}
The forthcoming sixth generation of wireless networks (6G) is expected to bring unprecedented transformations in terms of the developed wireless applications, use-cases, and the respective network requirements. On the one hand, it is forecast that an observable amount of sensing-centric and sensing-aided applications will represent a key pillar in the 6G, such as automatic driving, vehicle-to-everything (V2X), and extended reality (XR) \cite{survey6gnew}. On the other hand, standardizing entities have already forecast several expected enhancements with respect to the current 5G in terms of various key performance indicators (KPIs). Notably, this includes a communication reliability of $99.99999 \%$, a $10$ million/km$^2$ of network density, and a centimeter-level positioning accuracy \cite{survey6g}. Thus, such a massive KPI boost is challenged on the other side by the spectrum scarcity challenge, which restricts the considered network to abide by the existing frequency bands. In fact, the growth in the number of communication and sensing applications puts more pressure on the congested spectrum \cite{spectrumsc}. Therefore, developing innovative spectrum-efficient solutions has been a necessity to meet such targeted KPIs in futuristic networks.

Over decades of evolution on wireless standards and technologies, the wireless propagation channel has been a \textit{de facto} uncontrollable block in a communication system, mainly due to the presence of inevitable and stochastic losses, such as the small-scale fading, free-space path-loss (FSPL), and shadowing. Such phenomena are usually caused either by the long transmitter-receiver distance or by the absence of a line-of-sight (LoS) link due to the existence of obstacles in the propagation medium. Thus, this has been calling for developing transmission and/or reception schemes able to cope with the current channel conditions and improve signal detection. For instance, the well-known multiple-input multiple-output (MIMO) technology, and later its massive MIMO (mMIMO) upgrade, has been forming a key pillar in 5G networks, due to its great
potential in facilitating enhanced signal detection, spatial multiplexing, and spectral efficiency \cite{mmimo}.
Nevertheless, despite such achieved gains, MIMO can hardly turn an unfavorable channel, e.g., a non-LoS (NLoS) channel, into a favorable one due to the imposed spatial limitations of the antenna arrays at the transceivers. Therefore, there has been a critical need for developing innovative techniques that can not only preserve the different spectrum efficiency gains, but also reconfigure/alter the wireless channels to overcome some of the aforementioned challenging propagation scenarios.

In the efforts to circumvent the aforementioned wireless channel-related-challenges, the concept of \textit{}{channel reconfigurability} has been gaining a remarkable interest over the past few years. The overarching idea lies in considering the wireless channel a controllable block in the wireless system, rather than a black box. In this optic, \textit{reconfigurable intelligent surfaces (RIS)} was pioneered as a notable technique capable of changing the wireless channel response and overcoming the aforementioned challenges. RIS are manufactured planar surfaces composed of various reflecting elements (REs), each with a tunable electrical impedance that can control the phase shift and the amplitude of incoming signal waves to the respective RE, which controls its reflection angle and intensity. Thus, a proper adjustment of the REs' phase shifts can beamsteer the information signal to the users/area of interest with a considerably higher power by creating a constructive summation of reflected signal waves. RIS features various use cases in which it can circumvent wireless channel limitations. First, an RIS can create a \textit{virtual LoS link} to receivers facing the transmitter with an obstructed/absent direct link. Also, an RIS can provide steerable beams towards directions of interest by virtue of its tunable phase shifts, shaping the channel's array response. The increase in the RIS size can enhance the beamforming gain and compensate for the direct link's loss.

More recently, another technique was innovated in the context of reconfigurable channels, called \textit{movable antennas (MAs)}. The key idea of MAs stems from forming an antenna array without restrictions on its inter-element spacing or on its uniformity (e.g., equidistant spacing between elements). Thus, MAs are an array of antennas with tunable positions with mechanical controllers that can shift the location of each MA by a few wavelengths, either to its sides on the array axis (for linear arrays) or on both axes in a two-dimensional planar array \cite{movant1}. Thus, such a potential of repositioning the set of radiating antennas can offer additional degrees of freedom in modifying the channel response and improving the channel quality of the various users served. Such a position adjustment flexibility can result in an increase spectral efficiency compared to the typical uniform arrays \cite{movant2}.
Though the aforementioned two techniques have provided several gains in coverage and spectral efficiency, they exhibit several limitations when faced by the aforementioned channel-related challenges, i.e., 
\begin{itemize}
    \item On the one hand, despite MAs' flexibility in altering the channel properties, it cannot solve the absence of a LoS issue. MAs are typically movable over a scale of few (up to tens) wavelengths, rendering it unable to overcome an obstructed/absent LoS.
    \item On the other hand, while RIS can circumvent the LoS absence or obstruction by means of an indirect reflection, it essentially suffers from the double FSPL attenuation. Such an impairment is accentuated when operating with either a higher frequency band or when the transmitter-RIS or RIS-receiver distances are relatively long, which induces an observable trade-off between the RIS deployment cost and the achievable system performance.
\end{itemize}

\subsection{Pinching Antennas: Overview}

Recently, a promising alternative, known as pinching-antenna systems (PASS), has
been proposed and prototyped by NTT DOCOMO \cite{docomo}. The key principle in PASS lies in utilizing low-loss
long dielectric waveguides, extended from the transmitter and often installed in the facade of a building or in the ceiling of an indoor area, to guide signals and radiate them via small
dielectric elements, known as pinching antennas (PAs). The latter are positioned manually at custom points of interest along the waveguide. Accordingly, PASS presents various advantages compared to traditional MIMO, RIS, and MAs. First, the flexible positioning of antennas over a long waveguide (typically in the order of tens of meters) can help in overcoming the LoS blockage/absence issue for some users or coverage zones. That is, by pinching PAs at a convenient location in the waveguide with respect to the served user/area, a LoS link can be guaranteed. Furthermore, the placement of a PA close to the served user/area can ensure a lower FSPL attenuation, which results in a significantly higher received signal power. Such properties of PASS can pave the way for higher communication coverage by adopting longer cables that cover a wider zone \cite{mag1}.

Beyond communication reliability, coverage, and throughput enhancement, PASS exhibits additional benefits with respect to other concepts of RF propagation. PASS can enable sensing capabilities to track, localize, or detect unseen targets, e.g., obstructed by obstacles, by adequately placing transmitting and receiving PAs closer to the sensed targets with a clear LoS link \cite{mag2}. On another front, PASS can strike additional advantages in terms of information-theoretic security, i.e., physical-layer security (PLS). In PLS, one of the critical requirements to establish an information-theoretic secure transmission is having a legitimate channel superiority over the illegitimate one. Thus, traditional MIMO schemes can fail in ensuring such a channel advantage in some unfavorable channel settings. For instance, when only a dominant LoS component exists, an eavesdropper positioned at a similar azimuth and elevation angles as the legitimate user (with respect to the multi-antenna transmitter) yields an inevitable signal leakage as both receivers are positioned in the direction of the main signal beam. Therefore, the flexible antenna positioning in PASS can create a certain dissimilarity between both channels, which can serve in directing the legitimate signal to the intended (genuine) user by pinching beamforming.

\subsection{Related Surveys and Tutorials}

Despite the fact that the first successful demonstration of PASS by NTT DOCOMO dates back to 2022 \cite{docomo}, the first work witnessed on the design and analysis of robust schemes for PASS was not witnessed until the end of 2024 \cite{lit1}. As a result, only a handful of related surveys, tutorials, and short reviews have been documented so far on PASS. For instance, the work in \cite{mag1} and \cite{mag2} represents seminal reviews on PASS, presenting a technical high-level overview on their main principle, propagation model, and achievable gains in terms of coverage and throughput compared to traditional multi-antenna systems. While the latter work focused on presenting the underlying PASS architectures and a high-level presentation of PASS schemes on different applications, the former review presented a formal mathematical model for the in-waveguide and wireless propagation on PASS. In addition, the work of \cite{tutorial1} serves as a tutorial on PASS, presenting fundamentals of PASS in terms of in-waveguide electromagnetic propagation, signal modeling, and the various hardware models used for PASS deployment. The work discusses also the various antenna activation schemes, quantifies the 
performance gains of PASS, and gives an overview of model-, heuristic-, and ML-based PASS optimization techniques. Another related tutorial was reported in \cite{tutorial2}, whereby the authors present the concept of generalized PASS, encompassing different physical realizations of PASS. The work presented a holistic view on the single- and multi-waveguide-based PASS architectures, with a further categorization per the number of PAs deployed and the multiple-access scheme used, i.e., orthogonal and non-orthogonal. Closed-form expressions for the optimal PASS deployment are provided for simpler setups, while details on optimization approaches are given for more generalizing schemes. The survey of Liu \textit{et al.} in \cite{survey1} represents the first work to review the various contributions performed on PASS design, particularly on sensing-aided PASS. The work also introduces two additional waveguide architecture, namely the segmented and the multi-mode waveguide designs, in addition to the center-fed one. It should be noted that other surveys, such as \cite{survey2,survey3}, have covered the concept of reconfigurable/flexible antenna systems, however, they covered PASS as a particular case only. Nonetheless, their coverage did not touch upon critical use-cases in PASS, such as PASS for AI/machine learning (ML) systems, 
\subsection{Motivation and Contributions}
By reviewing the aforementioned survey and tutorial work provided in the body of research on PASS, their scope was limited in one or various aspects. First, it should be noted that the short reviews on PASS in \cite{mag1,mag2} were essentially to introduce the PASS concept and principles. Additionally, tutorials in \cite{tutorial1,tutorial2} were constrained to formalizing a mathematical framework for PASS analysis as well as presenting the various underlying architectures and optimization methods, and did not cover a thorough system-wise review of the work reported on PASS. Furthermore, the survey work in \cite{survey1} was dedicated to presenting the various works on PASS from a wireless sensing perspective. Lastly, despite tackling the PASS concept as a category of reconfigurable antennas in \cite{survey2,survey3}, a system-wise and thorough review of work done exclusively on PASS over the various architectures is missing.

Capitalizing on the above limitations of previous work, the current survey paper provides a thorough, comprehensive, and system-wise review of PASS, considering communication reliability, security, energy-efficiency, and sensing perspectives. In particular, the scope of work reviewed is categorized into schemes designed for several objectives or to assist various techniques, namely power efficiency, data rate, PLS, multiple-access, integrated sensing and communication (ISAC), and cutting-edge technologies. Such a system-wise review can help readers better situate their contribution to the existing body of research work done on PASS. In addition, we explore the potential of PASS in cutting-edge technologies, such as federated learning (FL), discussing the proposed PASS to circumvent some of the existing challenges of FL. Table \ref{complit} provides a high-level comparison of the current work's coverage compared to the aforementioned related surveys and tutorials.

\begin{figure}[!tbp]
\begin{center}
\includegraphics[scale=.59]{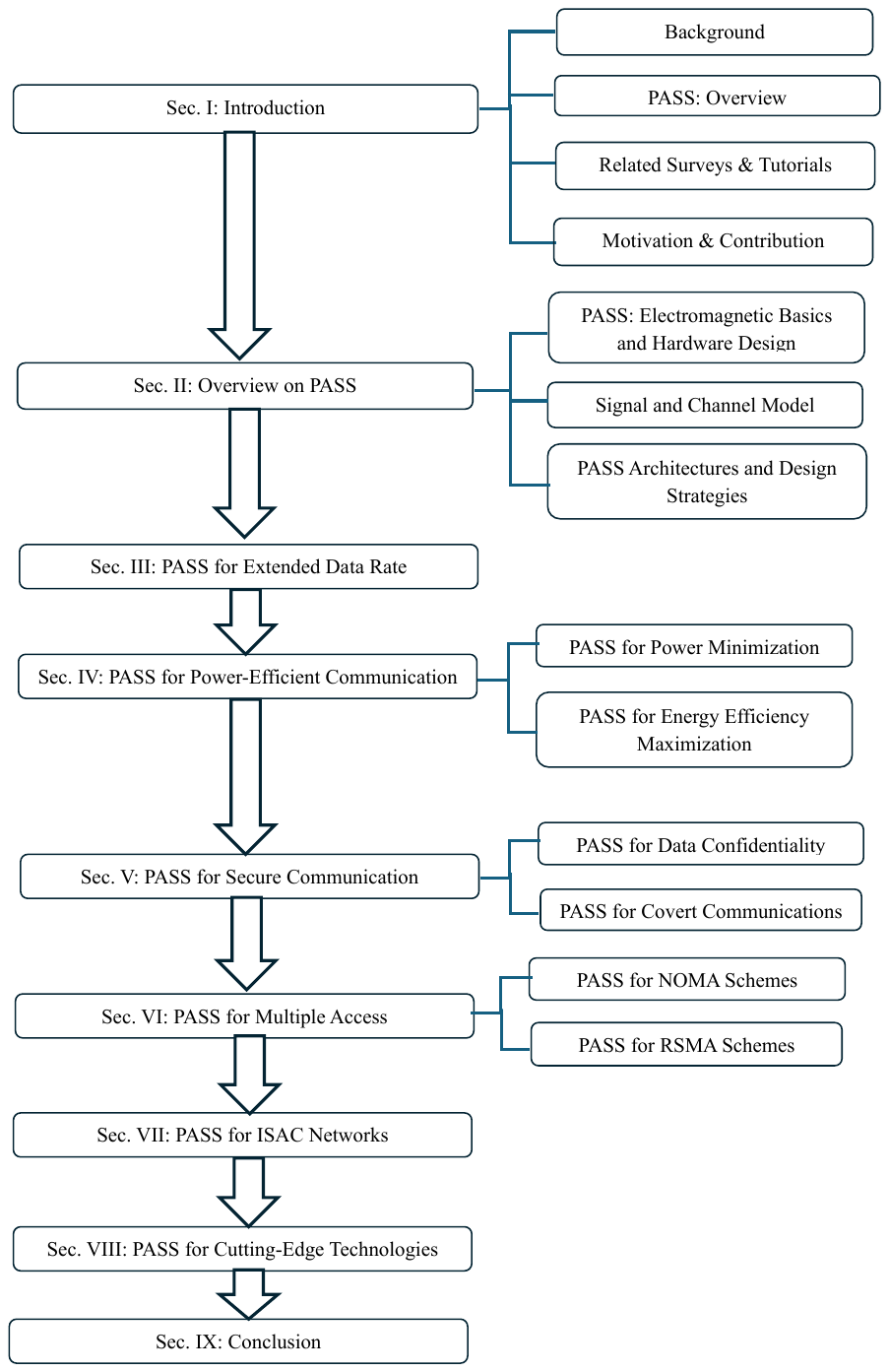}
\end{center}
\caption{\textcolor{black}{Paper organization.}}
\label{paperorg}
\end{figure}

As shown in Fig. \ref{paperorg}, the remainder of this paper is organized as follows: Section \ref{passoverview} presents a general background on the PASS concept, its various architectures, and enabling techniques from an electromagnetic perspective. Section \ref{passdatarate} reviews proposed PASS schemes to enhance data rate and coverage, while Section \ref{passpoweff} presents the work on PASS for realizing power-efficient wireless networks. PASS-aided secure wireless network designs are surveyed in Section \ref{passpls}, whereas Section \ref{passmultacc} discusses proposed PASS designs for different multiple access schemes. Then, Section \ref{passisac} presents the PASS designs for establishing ISAC networks, while Section \ref{passcutedge} discusses the proposed PASS schemes assisting FL-enabled networks. Existing and foreseen challenges related to implementing and deploying PASS are discussed thoroughly in Section \ref{passchallenges} with several future research perspectives, while Section \ref{passconclusion} concludes the paper.

\begin{table*}[t]

\renewcommand{\arraystretch}{2.3}
\centering
\phantom{~}\noindent

\begin{tabular}{|p{0.12\textwidth}|p{0.03\textwidth}|p{0.06\textwidth}|p{0.07\textwidth}|p{0.06\textwidth}
|p{0.06\textwidth}|p{0.06\textwidth}|p{0.09\textwidth}|p{0.04\textwidth}|p{0.04
\textwidth}|p{0.07\textwidth}|}

\hline\hline

\textbf{Work} & \textbf{Year} & {\textbf{Work} \newline \textbf{category}} & \textbf{Commun.} & \textbf{Sensing} & {\textbf{PLS (Confid.)}} & {\textbf{PLS (Covert.)}} & \textbf{PASS for Federated Learning} & \textbf{RSMA} & \textbf{NOMA} &  \textbf{Energy Efficiency}  \\
\hline\hline
Liu \textit{et al.} \cite{tutorial1} & 2025 & {Tutorial} & $\checkmark$ & $\star$ & $\star$ & \xmark & \xmark  & \xmark  & $\star$ & \xmark   \\
\hline
Xu \textit{et al.} \cite{tutorial2} & 2026 & Tutorial & $\checkmark$ & $\checkmark$ & $\star$ & $\star$ & \xmark & \xmark & $\checkmark$ & $\star$  \\ \hline
Liu \textit{et al.} \cite{survey1} & 2026 & Survey & \xmark & $\checkmark$ &  $\star$ & $\star$ & \xmark & $\star$ & $\star$ & $\star$ \ \\ \hline
\textit{Current work} & 2026 & $\checkmark$ & $\checkmark$ & $\checkmark$ & $\checkmark$ & $\checkmark$ & $\checkmark$ & $\checkmark$ & $\checkmark$ & $\checkmark$   \\ \hline
\end{tabular}
    \caption{\textcolor{black}{Comparison of related surveys and tutorials on PASS. $(\checkmark)$ refers to a well-covered part, ($\star$) indicates a limited coverage on a given part/concept, while (\xmark) means that the corresponding part was not covered.}}
    \label{complit}
\end{table*}

\section{Overview on Pinching Antenna Systems}
\label{passoverview}

In this section, a holistic overview of the functioning principles of PASS is presented. In particular, the section serves as a material for presenting the required background, from an electromagnetic propagation and mathematical modeling perspective of the PA-aided transmission and/or reception. The section also presents the existing models considered so far in the surveyed literature to account for the in-waveguide loss. 

\subsection{PASS: Electromagnetic Basics and Hardware Design}

A PASS is generally constituted of three main components, namely: 
\begin{itemize}
    \item A BS, representing the central processing unit of the system and equipped with radio-frequency (RF) chains and signal processing modules onboard.
    \item The dielectric waveguide, representing the guiding medium for the electromagnetic wave fed by the BS's RF chain,
    \item The PA, which is a small dielectric element pinched (attached) to the waveguide at a custom position to create a controlled signal radiation.
\end{itemize}

Dielectric waveguides are usually linked to the BS by virtue of wired connections (e.g., coaxial cable or optical fiber). Thus, dielectric waveguides represents a vital component of PASS. Their general architecture is composed of a core fabricated from dielectric material with a high refractive index with respect to its surrounding medium. Consequently, such a difference in refractive index can ensure the wave to be confined inside the core. The outer envelope of the core is often known as the cladding, which has a lower refractive index compared to the core \cite{refdielectric}. To this end, the overarching idea of PASS extends from the concept of leaky waveguides, which lies in creating an intentional and controlled energy leakage at certain positions of the waveguide \cite{tutorial2}. Herein, a PA can be mechanically placed at a position of interest to create a favorable radiation to serve a given user or area. 

\subsection{Signal and Channel Model}

\begin{figure}[h]
\par
\vspace*{-.3cm}\begin{center}
\includegraphics[scale=.30]{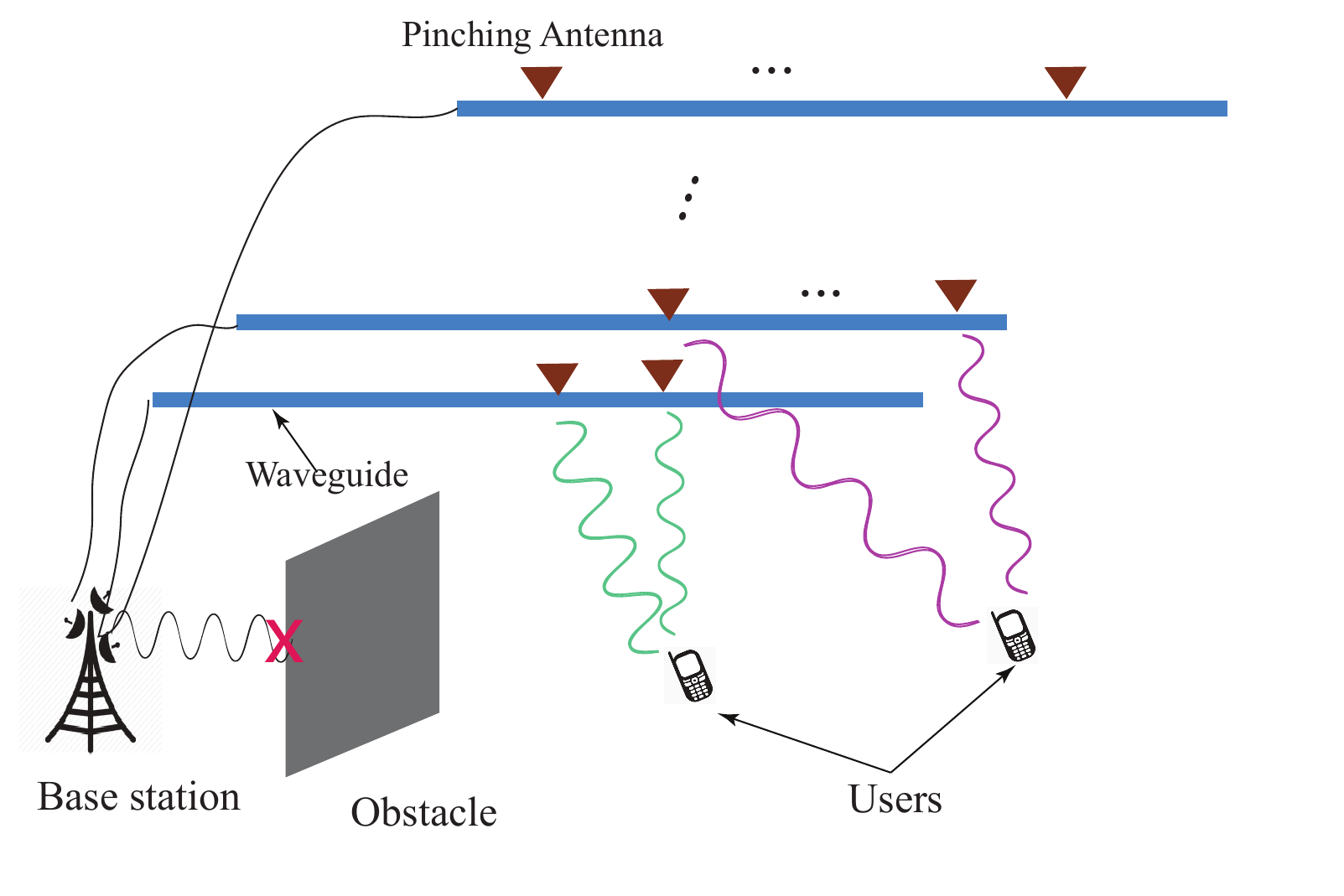}
\end{center}
\par
\captionsetup{font=footnotesize}
\caption{{Illustration of a multi-waveguide multi-PA PASS.}}
\label{pass1}
\end{figure}

\begin{figure}[h]
\par
\vspace*{-.3cm}\begin{center}
\includegraphics[scale=.30]{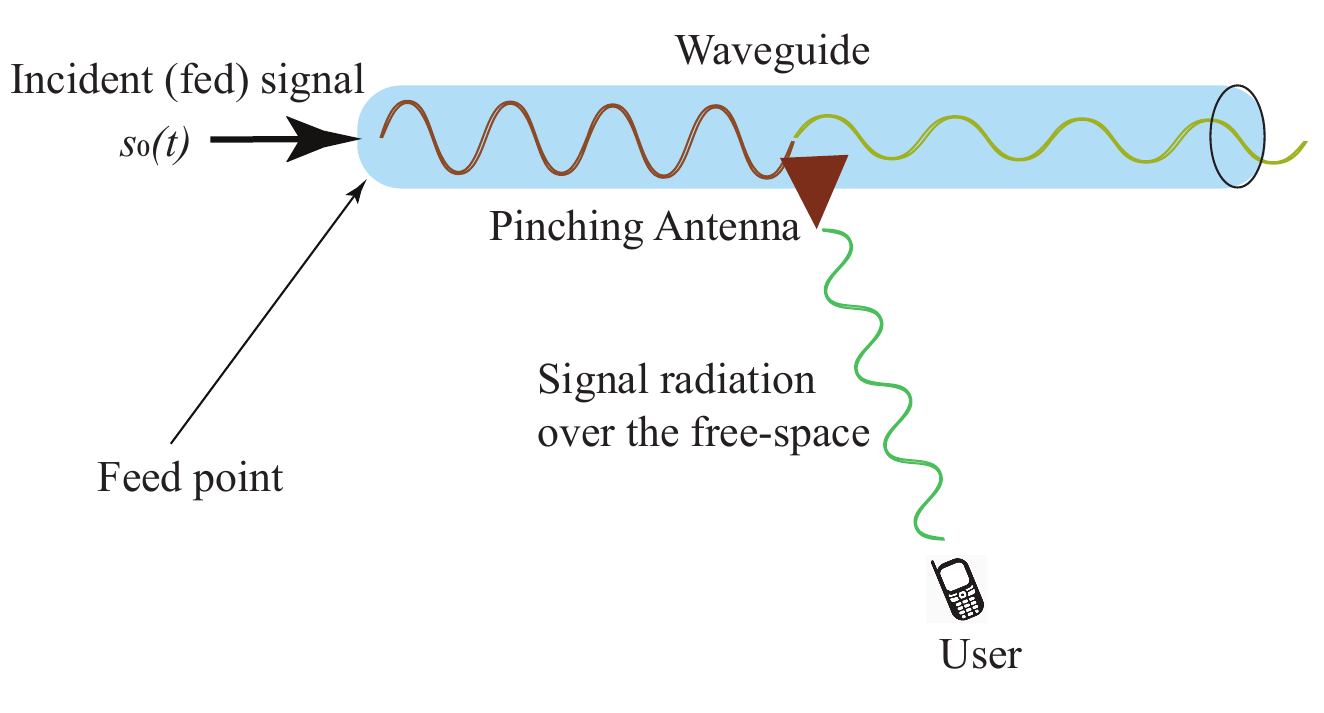}
\end{center}
\par
\captionsetup{font=footnotesize}
\caption{{Illustration of the in-waveguide and free-space propagation phase in a PASS.}}
\label{pass2}
\end{figure}

Let us consider a PASS illustrated in Fig. \ref{pass1}, consisting of a transmitting BS $(B)$ communicating with a receiver, denoted by  $(U)$. The PASS is based on extending $N$ waveguides from the BS in a parallel way. In the considered PASS, several PAs can be activated along a single waveguide. Let us consider that all the dielectric waveguides are parallel to the $x$-axis, where we denote by $\mathbf{p}_U=[x_U,y_U,z_U]$ and $\mathbf{p}_{B^{(m)}_{n}}=[x_{B^{(m)}_{n}},y_{B_{n}},z_{B_{n}}]$ as the location vectors for the served user and the $m$th PA of the $n$th waveguide, denoted by ${B^{(m)}_{n}}$. Observe that the $M$ PAs on the same waveguide share the same coordinate on the $y$ and $z$ axes, denoted by $y_{B_{n}}$ and $z_{B_{n}}$, respectively. To this end, the signal propagation in PASS can be decomposed into two stages, namely (i) the in-waveguide propagation and (ii) the free-space one, which will be detailed in the sequel. Fig. \ref{pass2} illustrates both aforementioned propagation phases.

\subsubsection{In-Waveguide Propagation}

As far as the in-waveguide propagation is concerned, a signal $s_0(t)$ is fed into the waveguide from a feeding point, denoted by $\mathbf{p}^{(\mathrm{feed})}_{n}=[0,y_{B_n},z_{B_n}]$. Inside the waveguide, the propagating wave is affected by (i) an attenuation caused by finite conductor resistance and dielectric loss. Thus, one can link the signal inside the waveguide after traveling a distance $L$ and the injected signal $s_0(t)$ as \cite{survey1}
\begin{equation}
    s_L(t)=\sqrt{P} \exp(-\delta L) s_0(t) \label{rxsig}
\end{equation}
where $P$ is the transmit signal power, $\delta=\alpha+j\beta$ is defined as the in-waveguide attenuation coefficient in m$^{-1}$. Note that the complex-valued attenuation coefficient is decomposed into an amplitude attenuation term $(\alpha)$ and phase shift-inducing coefficient $(\beta)$. The latter is defined as $\beta \triangleq 2 \pi n_W/\lambda $ where $n_W$ is the in-waveguide refractive index and $\lambda$ is the operating wavelength. A lossless waveguide yields $\alpha=0$, resulting in a signal that is only phase-shifted along the waveguide. For a single-input single-output (SISO) PASS with a single waveguide and $M$ PAs, the signal radiated from the $m$th PA can be expressed as 
\begin{align}
    s^{(m)}_{\mathrm{pin}}(t) & = \sqrt{P_m G_{m,n}^{(\mathrm{T})}(\theta,\phi)} \exp(-\alpha \Vert \mathbf{p}_{B^{(m)}_{n}} -\mathbf{p}^{(\mathrm{feed})}_{n}\Vert_2) \notag \\ & \times \exp(-j {2\pi n_W \Vert \mathbf{p}_{B^{(m)}_{n}} -\mathbf{p}^{(\mathrm{feed})}_{n}\Vert_2}/{\lambda}) s_0(t)\label{rxsig2}
\end{align}

In \eqref{rxsig2}, $\Vert . \Vert_2$ defines the L$2$-norm, and the term $\Vert \mathbf{p}_{B^{(m)}_{n}} -\mathbf{p}^{(\mathrm{feed})}_{n}\Vert_2$ stands for the distance that the signal traveled on the waveguide before being radiated by the $m$th PA. It can be easily found that $\Vert \mathbf{p}_{B^{(m)}_{n}} -\mathbf{p}^{(\mathrm{feed})}_{n}\Vert_2=x_{B^{(m)}_{n}}$. In addition, $P_m$ denotes the portion of power radiated by the PA, and $G_{m,n}^{(\mathrm{T})}$ defines the radiation gain of the transmit PA, given in terms of the azimuth and elevation angles. Generally, for a signal of power $P$ supplied at to the waveguide, the pinching power (i.e., radiated power from a PA) is $P_{\mathrm{pinch}} = P \sin ^2 (\kappa L)$ where $\kappa$ is the mode coupling coefficient, and $L$ is the length of the PA. On the other hand, the remaining power at the waveguide after pinching at the $m$th PA is $P_{\mathrm{rem}} = P \cos^2 (\kappa L)$. Accordingly, for an $M$-antenna waveguide, the radiation power of the $m$th PA can be formulated as \cite[Eq. (19)]{passpowmin1}
\begin{equation}
    P_m= P \delta_m^2 \prod_{l=1}^{m-1} \sqrt{1-\delta_l^2}
    \label{pinchpowm}
\end{equation}
where $\delta_m \triangleq \sin (\kappa L_m)$ and $L_m$ is the length of the $m$th PA. Observe from \eqref{pinchpowm} the dependence of the radiated power at the $m$th PA on the portion of remaining power, which depends per se on the power radiated by preceding antennas. Also, one can note that $P_m$ in \eqref{pinchpowm} is written as the product of two terms, namely the ratio of power radiated from the PA, i.e., $\delta_m$, as well as the total power remaining at the waveguide, i.e., $ \prod_{l=1}^{m-1} \sqrt{1-\delta_l}$. Notice that $\delta_i$ $(\forall i)$ depends on the length of the $i$th PA, which, in turn, controls the amount of power radiated. Therefore, there are two different power distribution modes distinguished, namely:
\begin{itemize}
    \item \textbf{Equal Power Model}: Herein, the amount of radiated power for all the antennas is the same. Thus, PAs are manufactured with different lengths to allow for radiating the same power $P_m=P_{\mathrm{eq}}$ $(\forall m)$. Thus, such a model allows essentially for ensuring a radiation power \textbf{fairness} among the various antennas, irrespective of their position in the waveguide. Such a model is useful for obtaining insights into analytical performance evaluation.
    \item \textbf{Proportional Power Model}:  In this model, it is assumed that all PAs are manufactured with the same length, $L_m = L \Leftrightarrow  \delta_m=\delta$ $(\forall m)$. Thus, the amount of power radiated differs from one PA to another, i.e., $P_i \neq P_j$, as noted from \eqref{pinchpowm} and the definition of $\delta_m$. However, such a model allows for a less complex manufacturing design since the set of PAs are manufactured identically.
\end{itemize} 

\subsubsection{Free-Space Propagation}

Once the electromagnetic wave is radiated by a given PA, free-space wireless propagation take place, which exhibits the same phenomena as in the traditional multi-antenna systems, i.e., FSPL, small-scale fading, and shadowing. Thus, one can represent the received signal at a given receiver $U$, located in a position $\mathbf{p}_U$, from $B_{n}^{(m)}$ as
\begin{equation}
    y_U(t)=\frac{\lambda \sqrt{G_R(\theta,\phi)}}{4 \pi \Vert \mathbf{p}_{B^{(m)}_{n}}-\mathbf{p}_U\Vert_2} h_{m,n,U}  s^{(m)}_{\mathrm{pin}}(t)+w_U \label{rxsig3}
\end{equation}
with $G_R(\theta,\phi)$ is the receive antenna gain at $U$,
\begin{align}
    h_{m,n,U}&=\sqrt{K/(K+1)}\exp(-j2\pi/\lambda \Vert\mathbf{p}_{B^{(m)}_{n}}-\mathbf{p}_U\Vert_2)  \notag \\ & + \sqrt{1/(K+1)}  h^{\mathrm{(NLoS)}}_{m,n,U}
\end{align} is the small-scale fading coefficient for the link between $B^{(m)}_{n}$ and $U$, composed of a LoS link (first term) and an NLoS one (second term). Note that $\Vert \mathbf{p}_{B^{(m)}_{n}}-\mathbf{p}_U\Vert_2$ represents the distance between $B^{(m)}_{n}$ and $U$. Additionally, $K$ is the Rician-$K$ factor, and $w_U$ is the additive white Gaussian noise at $U$. 

\subsubsection{Extension to Multiple Waveguides}
For a multi-PA multi-waveguide PASS, $U$ receives various signal copies from the set of PAs pinched along the $N$ waveguides. Herein, each waveguide is ideally connected to a dedicated RF chain. On the one hand, an identical signal is fed to the set of PAs pinched along the same waveguide. Nonetheless, as observed from \eqref{rxsig} and \eqref{pinchpowm}, the power and phase shift of the radiated version by each PA depends on (i) its location and (ii) the number of PAs preceding it. On the other hand, a distinct signal is fed to each waveguide. Such a setup can create a similar setup to a conventional multi-antenna transmitter, where the signal fed to each waveguide is precoded by a certain weight $f_n$ $(n=1,\ldots,N)$ in order to achieve spatial multiplexing and beamforming gains. Consequently, one can express the received signal at $U$, based on \eqref{rxsig2}-\eqref{rxsig3} as
\begin{equation}
    y_U(t)=\sum_{n=1}^{N}\mathbf{h}_{B_{n}U}\left( \mathbf{p}_{B_n}\right) \mathbf{g
}\left( \mathbf{p}_{B_n}\right) s_{n}(t)+w_{U}
\end{equation}
where 
\begin{equation}
    \mathbf{h}_{B_{k}U} \triangleq \left[ \eta_{1,n}h_{1,n,U} , \ldots ,   \eta_{M,n} h_{M,n,U} \right] \in \mathbb{C}^{1\times M}
\end{equation}
 $ \eta_{m,n} \triangleq {\lambda \sqrt{G_R(\theta,\phi)}}/{(4 \pi \Vert \mathbf{p}_{B^{(m)}_{n}}-\mathbf{p}_U\Vert_2)}$, $ \mathbf{p}_{B_n} \triangleq  \left[ \mathbf{p}_{B_n^{(1)}}, \ldots, \mathbf{p}_{B_n^{(M)}}\right]$,
\begin{equation}
    \mathbf{g
}\left( \mathbf{p}_{B_n}\right)  \triangleq \left[ \nu_{1,n}g_{1,n} , \ldots ,   \nu_{M,n} g_{M,n} \right]^T \in \mathbb{C}^{M\times 1},
\end{equation}
\begin{equation}
    \nu_{m,n} \triangleq \sqrt{P_m G_{m,n}^{(\mathrm{T})}(\theta,\phi)} ,
\end{equation}
and 
\begin{equation}
    g_{m,n}=\exp(-(\alpha+j {2\pi n_W /{\lambda})\Vert \mathbf{p}_{B^{(m)}_{n}} -\mathbf{p}^{(\mathrm{feed})}_{n}\Vert_2}) .
\end{equation}

\subsection{PASS Architectures and Design Strategies}

After detailing the fundamental channel and signal propagation models in PASS, an overview of enabling architectures is presented and discussed in this subsection. 

\subsubsection{Alternative PASS Medium Design}
Apart from the well-known dielectric waveguide-based design, PASS realization can be established with alternative designs, mainly based on leaky coaxial cables (LCX) as well as pinching-inspired antenna arrays. 

\paragraph{Leaky Coaxial Cables}

As far as LCXs are concerned, their operation principle differs from the waveguide-enabled prototype designed in \cite{docomo} in the controllable radiation aspect. In particular, LCXs radiate signals by virtue of periodic slots set along the cable. Thus, such a design lacks full control of the PA positioning flexibility since the radiation leakage points cannot be reconfigured \cite{lcx}. Nonetheless, an adjustable design of LCXs-based PASS can be established by dividing an LCX into various segments, each of which is controlled by an electronic switch to control the segments (sections) that radiate the information signal \cite{tutorial2}. 

\paragraph{Pinching-Inspired Antenna Arrays}

Another architecture that can be utilized for realizing PASS is the pinching-inspired antenna arrays. Herein, the electromagnetic wave is guided through a metallic tube using the surface-wave mode. Then, a radiating element (antenna) can be positioned in a position of interest, e.g., close to a user or coverage zone. A particular example of interest is the radio stripes concept utilized in cell-free massive MIMO \cite{radiostripes}.

\subsubsection{Promising Architectures}

Conventional PASS designs are based on linking one or many waveguides to the BS by virtue of dedicated wired connections, where each waveguide is linked to a dedicated RF chain. In this optic, some promising architectures, extending the traditional one, have been recently proposed, which we will present in the sequel.

\paragraph{Segmented Waveguide-Enabled PASS}

Despite the provable gains of waveguide-enabled PASS in terms of coverage and data rate, some challenges can limit its performance, namely
\begin{itemize}
    \item \textbf{In-Waveguide Attenuation:} While many studies overlooked the presence of the in-waveguide attenuation coefficient, empirical analysis demonstrates a dependence of such an attenuation coefficient on the waveguide structure and the operating frequency.
    \item \textbf{Inter-Antenna Radiation:} For uplink PASS, the received signal at each PA pinched on the waveguide may be leaked through other PAs on its way to the BS. 
\end{itemize}

\begin{figure}[h]
\par
\vspace*{-.3cm}\begin{center}
\includegraphics[scale=.43]{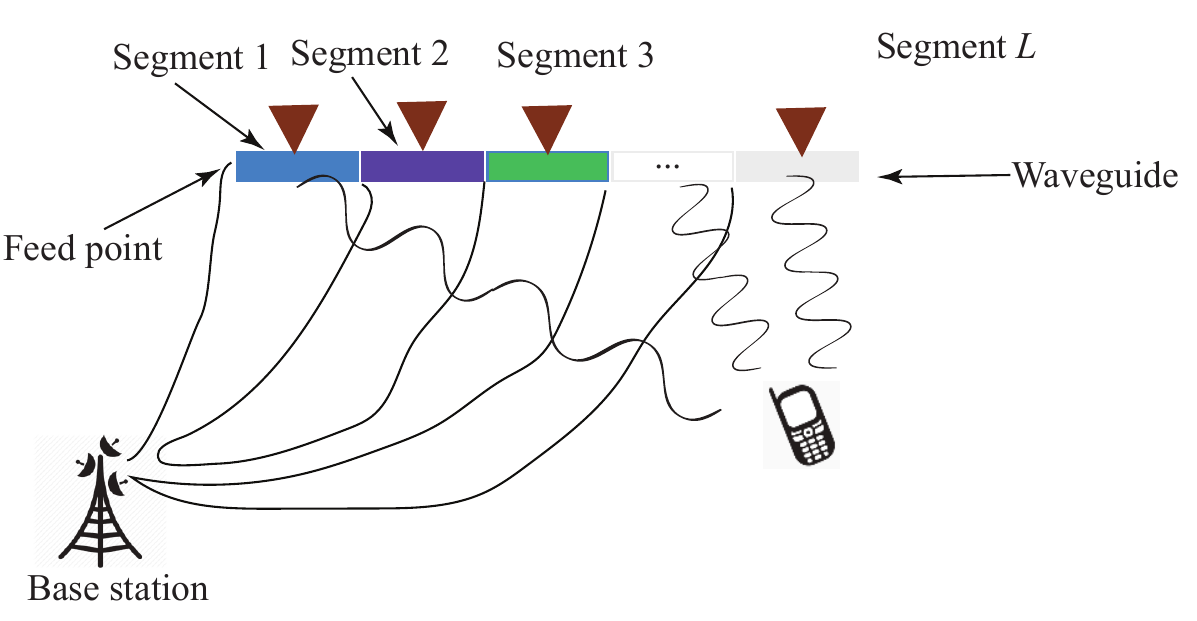}
\end{center}
\par
\captionsetup{font=footnotesize}
\caption{Illustration of a Segmented Waveguide Pinching Antenna (SWAN) System.}
\label{swan}
\end{figure}

Thus, as a remedy to the above challenges, a novel PASS architecture, called segmented waveguide-enabled pinching antenna (SWAN), was proposed in \cite{passrate3}. As shown in Fig. \ref{swan}, SWAN considers each waveguide to be divided into different segments, each with a dedicated feed point. Such a design can essentially help in (i) reducing the in-waveguide loss by reducing the propagation distance from the feed point to radiation points and (ii) reducing the inter-antenna radiation effect. Thus, the segments are isolated, and each segment's feed point is linked with an independent RF chain. A comprehensive analysis is performed on three different SWAN PASS designs, namely segment selection (SS), segment aggregation (SA), and segment multiplexing (SM). In SS, only a single segment is connected to the single RF chain operating, while SA is based on connecting all the segments to the single RF chain. SA can achieve a better performance compared to SS by connecting all the segments to the RF chain by virtue of a power splitter. On the other hand, SM stands out as the best strategy among the three SWAN strategies due to the dedicated RF chain per waveguide segment, which enables distinct signals to propagate on each segment and, consequently, the implementation of spatial multiplexing and beamforming schemes.

\paragraph{Center-Fed PASS}

\begin{figure}[h]
\par
\vspace*{-.3cm}\begin{center}
\includegraphics[scale=.33]{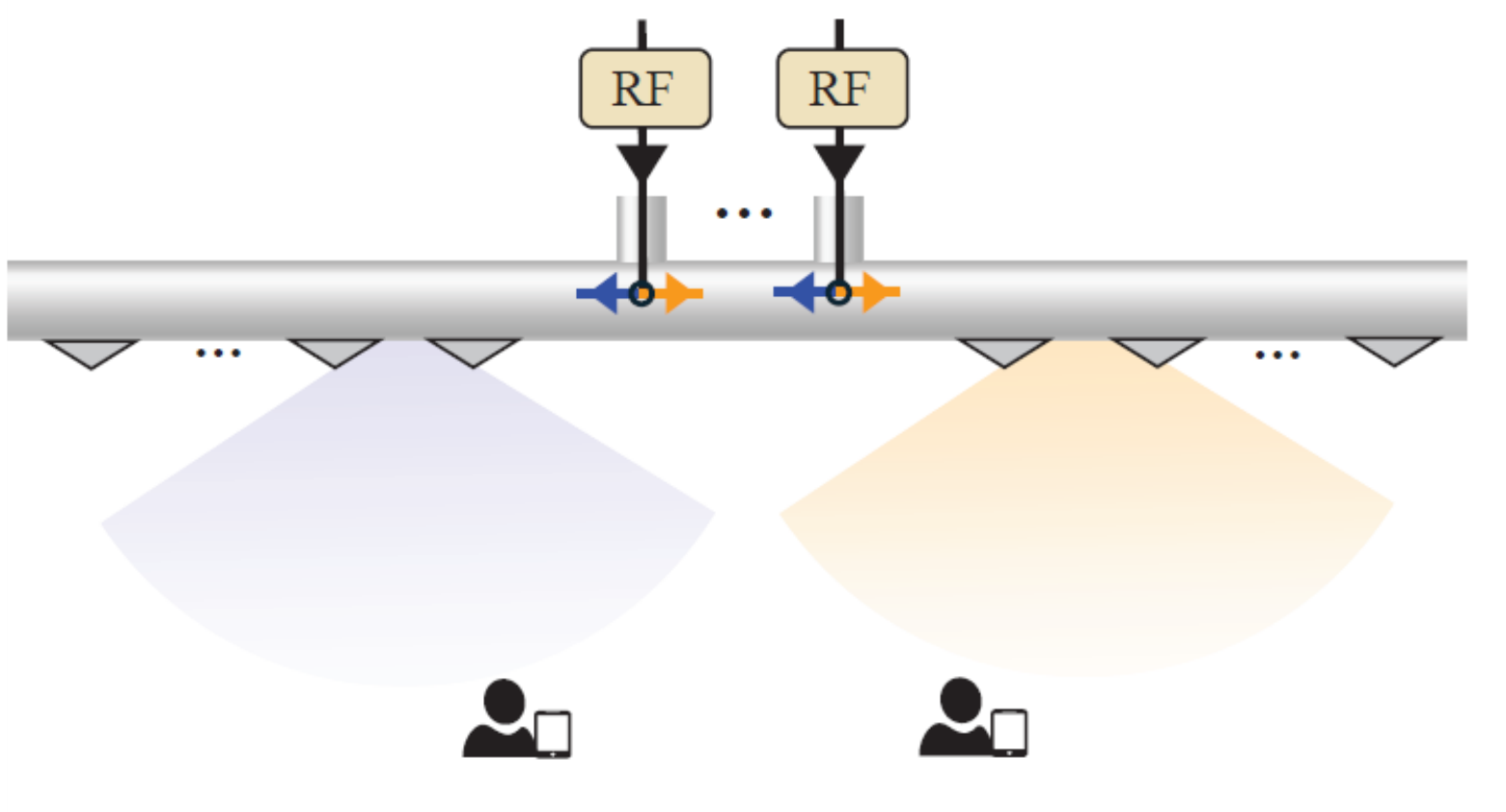}
\end{center}
\par
\captionsetup{font=footnotesize}
\caption{Illustration of a Center-Fed PASS (C-PASS) \cite{cpass}.}
\label{cpassfig}
\end{figure}

While multi-waveguide schemes achieve considerable data rate enhancements, their main challenges lie in the need to deploy multiple parallel waveguides with a sufficient separation to ensure non-coupling. On the other hand, single-waveguide PASS is limited in terms of its data rate, since only one signal can be fed into the waveguide. Thus, in a single-waveguide setup, users should share the existing resources (e.g., time slot, subcarriers). Thus, as an effort to increase PASS degrees of freedom, another enabling architecture was proposed, known as the center-fed PASS (C-PASS), given in Fig. \ref{cpassfig}. The basic principle of C-PASS lies in feeding the waveguide with one or many signals from input ports positioned anywhere in the waveguide except the waveguide's ends. Each signal is split in some way to let a portion of it propagate in the forward direction of the waveguide, while the remaining portion propagates in the backward direction. Thus, a combination of multiple signals fed from various points, split in a precise way, gives additional degrees of freedom to serve an additional user/area by the same waveguide, hence doubling the degrees of freedom per waveguide \cite{cpass,cpass2}.

The splitting process in C-PASS can be performed in three different strategies, as given in \cite{cpass}, as:
\begin{itemize}
    \item \textbf{Power Splitting}: In this scheme, a power splitter is utilized at each input port to split the power of the input signal between the forward and backward directions.

    \item \textbf{Direction Switching}: Herein, the input ports are divided to two groups, namely the ones directing the input signals to the forward directions and another portion of ports operating in the backward direction.

    \item \textbf{Time Switching}: Such a strategy is based on equipping each input port with a switch, enabling the switching of all the input ports to a given direction for a particular time slot.
\end{itemize}

\paragraph{Wireless-Fed PASS}

One of the main challenges of PASS is the difficulty to deploy waveguides in some environments, in addition to the in-waveguide attenuation when deploying a sufficiently long waveguide over a large area. In this optic, another architecture was proposed, known as wireless-fed PASS (Wi-PASS) \cite{wipass,wipass2}. In such a scheme, the backhaul link between the BS and a waveguide is established through a wireless link, where a relay, connected to the waveguide, receives, processes, and feeds the waveguide with the information signal. Such a design enables a flexible waveguide deployment, since the relay can be freely positioned near the coverage area. To this end, the waveguide can span only over a limited range of interest. Furthermore, the wired backhaul links between the BS and the waveguides can be eliminated in Wi-PASS.

\section{Pinching Antenna Systems for Extended Coverage and Higher Data Rate}
\label{passdatarate}

The work in \cite{lit1} is considered the seminal work on PASS, where the authors proposed a mathematical model for the received signal and channel in PASS, based on which a closed-form expression for the ergodic sum-rate of a multi-user PASS was derived. A closed-form expression was proposed for a single-waveguide setup, while a search-based optimization algorithm was proposed for an optimal PA placement in multi-waveguide schemes. The authors of \cite{lit3} analyzed a downlink single-waveguide PASS serving a single user. To maximize the downlink communication rate, a low-complexity two-stage optimization approach was developed. First, an initial positioning of the set of PAs is performed in order to minimize the FSPL. Then, the locations are further refined to create a coherent (constructive) aggregation of signals broadcast from the various PAs to the user. In \cite{lit4}, a PASS scheme was developed to maximize the sum rate of a multi-user downlink network. The considered PASS scheme upgrades the ones in \cite{lit1,lit3} by considering multiple waveguides. Thus, a transmit beamforming scheme can be established by broadcasting a distinct signal from each waveguide's PAs, scaled by a precoding weight. An alternating optimization (AO)-based framework was developed by alternatively optimizing the transmit beamforming vector and the PA positions using, respectively, a regularized zero-forcing (ZF) precoder and a grid search. 

While the aforementioned work relied on iterative methods (e.g., AO, successive convex approximation (SCA)) to convexify the sum rate or per-user rate in terms of the PA locations, assessing the quality of the sub-optimal solution is challenging. In this optic, learning-based methods stands out to better model and learn the gradient of the objective function, which helps in refining the optimal PA locations. Thus, the authors of \cite{lit6} proposed a graph neural network (GNN)-based framework to optimize the function of a downlink multi-waveguide multi-user PASS. The proposed data-based GNN framework optimizes the PA locations as well as the beamforming matrix to maximize the network's sum rate.

On the other hand, the work in \cite{lit7,lit8} conducted
a similar analysis and optimization of an uplink PASS. In particular, the authors of \cite{lit7} considered an uplink single-waveguide PASS aiming to serve various uplink users. By considering an orthogonal multiple access (OMA)-based scheme, each user can be allocated orthogonal time or frequency resources to avoid the inter-user interference at the receiver. With the aim of maximizing the minimum data rate among users, an AO-based algorithm is developed to optimize the PA positions that balance the FSPL and the constructive combining of signals, while the optimal transmit power levels are determined by a Lagrangian and SCA-based scheme. The work in \cite{lit8}extends the PASS scheme in \cite{lit7} by adopting a multi-waveguide setup to serve multiples uplink users. A sum rate maximization problem was formulated, for which an AO and gradient descent (GD)-based technique is proposed to sequentially optimize the PA location in each waveguide.

The authors in \cite{passrate1} studied a downlink MU multiple-input single-output (MISO) PASS, featuring two architectures, namely (i) a single-waveguide multicast scheme and (ii) a multi-group multicast one utilizing multiple waveguides operating with transmit beamforming. A CEO approach was utilized to optimize the location of the PAs for both architectures, which is based on iteratively (i) drawing a set of candidate solutions from a given distribution (e.g., multivariate Gaussian), followed by (ii) updating the sampling distribution mean and variance vectors minimizing the Kullback-Liebler divergence with a distribution built upon the drawn samples. 

Recently, Ouyang \textit{et al.} introduced in \cite{passrate3} a SWAN-enabled PASS. The three different SWAN PASS designs, namely SS, SM, and SA, were evaluated on both uplink and downlink single-user transmission. While SM results in the highest data rate performance among the three designs, its operating costs is relatively higher as it requires an RF chain for each segment to process, i.e., precode (combine) the transmitted (received) signal. 

The above-discussed work analyzed the data rate enhancement of PASS considering an infinite blocklength regime, which results in the well-known Shannon capacity (e.g., transmission rate achieving an arbitrarily small decoding error). Nonetheless, the current 5G involves also a short-packet class of services, known as the ultra-reliability low-latency communication (URLLC), focusing on achieving a reliable transmission of short packets. Thus, the blocklength becomes finite in such a case. In \cite{passrate2}, the authors analyzed the downlink rate enhancement of a single user PASS-aided network adopting an URLLC. A two-step PA placement procedure is designed to initially position the set of PAs close to the user, while further position finetuning is performed to ensure a coherent phase difference between the signals emanating from each pair of adjacent PAs. Though the above work investigated data rate improvements of PASS with respect to conventional multi-antenna systems, the underlying optimization involved dealing with complex optimization problems, which are usually solved using iterative procedures. This yields either suboptimal solutions of high complexity. As a remedy to these challenges, the work \cite{passrate4} explored the potential of deep learning (DL) in establishing a data-guided optimization of a PASS parameters. Particularly, the proposed approach leveraged a graph neural
networks (GNNs)- and multi-layer perceptron (MLP)-based scheme to select the optimal set of active PAs in a two-state single-waveguide PASS, with the objective to maximize the downlink rate of a served user. A more generalized PASS setup was investigated in \cite{passrate5}, where the authors considered a multi-waveguide PASS serving various users. Herein, a gradient-based meta-learning (GML) joint optimization algorithm was proposed to tackle the original non-convex problem. By virtue of convex altenative representation of the primal problem, two convex subproblems are distinguished, optimizing the beamforming vectors and PA locations. Two neural networks are used to solve the two aforementioned subproblems. The GML-based approach shows a remarkable gain in learning the gradient behavior and the required update step, and manifests a better resilience against the initial value of the solution or the used channel realization. Thus, an observable performance gain was noted compared to its AO counterpart. Additionally, the authors of \cite{passrate6} proposed a model-based and learning-based algorithms for maximizing the sum rate of a multi-user multi-waveguide PASS. While the first scheme consisted of a joint majorization minimization (MM) and penalty dual decomposition iterative approach, a Karush-Kuhn Tucker (KKT)-guided dual Lagrangian (KDL) approach is developed as a learning-based approach, considering a sequence of neural networks (NNs), i.e., deep unfolding-based optimization, to predict optimal the PA locations and beamforming matrices. Another similar multi-user multi-waveguide PASS setup was analyzed in \cite{passrate7}. Herein, a Transformer learner was adopted for optimizing analog and digital beamforming matrices, in addition to PA positions. The proposed Transformer learner adopts a graph-based architecture with permutation properties. Thus, unlike traditional learners in \cite{passrate5,passrate6}, the graph-based architecture ensures a scalability to any number of PAs and waveguides without an additional retraining cost. Thus, such an approach manifests better adaptability to PASS with respect to the change in the adopted setup.

On another front, while the above-detailed work provided several closed-form or algorithm-based solutions for the optimal PA placement, the analysis overlooked three core considerations:
\begin{itemize}
    \item The in-waveguide attenuation, which can reach up to $1.3$ dB/m \cite{docomo}.

    \item The PA directivity, which contradicts the usual assumption of an isotropic radiation by each PA.

    \item The probabilistic LoS blockage caused by the highly directional beams, particularly in mmWave bands.
\end{itemize}

To address such an analysis inconsistency, the authors in \cite{passdirectional} conducted a comprehensive analysis of a directional PASS (DiPASS), which considers the aforementioned three assumptions. The existence of an in-waveguide attenuation strikes a trade-off between coverage and signal power loss. With the objective of maximizing the sum rate of a multi-user multi-waveguide network, an optimization problem was formulated, including the optimization of the PAs' locations and orientation angles, in addition to the precoding matrix. An analytical evaluation for a single-PA single-user particular case revealed the existence of an optimal PA location, depending on the in-waveguide and FSPL attenuation exponents. 

The aforementioned work analyzed PASS' potential, essentially in circumventing FSPL attenuation. However, they considered that the LoS link is present in a PASS setup. However, despite the flexibility of antenna positioning in a PASS, dynamically present obstacles can block signal propagation from a PA to a served user. Aiming to investigate the impact of LoS blockage caused by obstacles on PASS, Ding et al. carried out in  \cite{passblockage2} a comprehensive analysis of a PASS with a probabilistic blockage model. By analyzing two designs, consisting namely of a single and multiple waveguides, remarkable gains in terms of the outage probability and ergodic data rate were observed with respect to a conventional multi-antenna system. Such results demonstrated a better resilience of a PASS to a LoS blockage compared to a conventional setup. The authors of \cite{passblockage} considered a generalized blockage-aware PASS framework. In particular, a deterministic 
LoS blockage model was adopted in the presence of cylinder-shaped obstacles. While LoS blockage can decrease the received signal power in PASS compared to a non-blocked LoS case, it can help in reducing the inter-user interference. Precisely, each PA can be positioned with a LoS link only to a particular user, while facing an obstructed link to other users. Thus, two schemes were proposed to leverage the potential of PASS in a blockage-aware environment, when either each antenna serves a single user or all users. While a Hungarian algorithm and a block coordinate descent (BCD)-based approach was proposed for the former scheme, a hybrid algorithmic approach was adopted for the latter, combining a weighted minimum mean-squared error (WMMSE) precoder and a deep reinforcement learning (DRL) algorithm for PA positioning. 
Table \ref{tabledatarate} summarizes the above-surveyed work.

\begin{table*}[t]
\centering
\caption{Summary of PASS-related designs for data rate and coverage enhancement.}
\label{tabledatarate}
\begin{tabular}{|p{3cm}|p{8cm}|p{4cm}|}
\hline
\textbf{Work} & \textbf{Contribution} & \textbf{Optimization Technique} \\ \hline

Ding et al. \cite{lit1} 
& Closed-form expression for the ergodic sum-rate of multi-user downlink PASS systems. Closed-form analysis for single-waveguide setups, while optimal PA placement and precoding were proposed for multi-waveguide schemes. 
& Search-based optimization \\ \hline

Xu et al. \cite{lit3} 
& Studied a downlink single-waveguide PASS serving a single user and maximized the communication rate. A two-stage placement method was proposed to first minimize FSPL and then refine PA locations for constructive signal combining. 
& Two-stage optimization (heuristic placement refinement) \\ \hline

Bereyhi et al. \cite{lit4} 
& Developed a multi-waveguide PASS scheme for multi-user downlink networks to maximize the sum rate. A transmit beamforming framework was designed with precoding weights across waveguides. 
& AO with ZF precoding and grid search \\ \hline

Guo et al. \cite{lit6} 
& A data-driven framework for multi-user multi-waveguide PASS using GNN to maximize the network sum rate.
& Graph Neural Network (GNN)-based learning \\ \hline

Tegos et al. \cite{lit7} 
& Investigated an uplink single-waveguide PASS with OMA to serve multiple users. The objective was maximizing the minimum user rate by jointly optimizing PA placement and user transmit power. 
& Alternating optimization, Lagrangian method of multipliers, and SCA \\ \hline

Zhang et al. \cite{lit8} 
& Extended uplink PASS to multi-waveguide architectures and formulated a sum-rate maximization problem for multi-user uplink transmission. 
& AO and GD \\ \hline

Chen et al. \cite{passrate1} 
& Studied a MU-MISO PASS system supporting single- and multi-waveguide multicast architectures.
& Cross-Entropy Optimization \\ \hline

Ouyang et al. \cite{passrate3} 
& Proposed the SWAN architecture for uplink and downlink PASS to mitigate in-waveguide losses and antenna coupling. Three segment strategies (SS, SA, SM) were analyzed in terms of achievable data rate. 
& Analytical evaluation \\ \hline

Lin et al. \cite{passrate2} 
& Investigated PASS-assisted URLLC transmission under finite blocklength. A two-step PA placement strategy was proposed to enhance reliability and achievable rate for short packets. 
& Two-step heuristic-based optimization (i.e., FSPL minimization followed by constructive signal summation)\\ \hline

Zhang et al. \cite{passdirectional} 
& Proposed a directional PASS model considering in-waveguide attenuation, PA directivity, and probabilistic LoS blockage. An optimization problem was formulated to maximize the multi-user sum rate by jointly optimizing PA locations, orientations, and precoding, considering the aforementioned limitations.
& Analytical study (single-PA) and iterative optimization (multi-PA) \\ \hline

\end{tabular}
\end{table*}

\subsection{Lessons Learned}

While the aforementioned work demonstrated notable gains of PASS in terms of achievable rate and flexibility, several key insights can be noted from the literature review that are critical for guiding future research and practical deployment. These observations highlight the trade-offs and modeling considerations that must be carefully addressed to optimize performance in realistic scenarios. In particular, the following points are highlighted:
\begin{itemize}
    \item It is crucial to assess the rate–complexity trade-off for the various optimization and design techniques used, including learning-based algorithms, AO-based schemes, and heuristic approaches. Each method offers varying computational overhead and convergence behavior, which directly impacts the practical feasibility and efficiency of PASS implementation.

    \item The use of a segmented waveguide architecture provides a promising avenue for compactly deploying spatial multiplexing schemes. As reported in \cite{passrate3}, SM typically achieves superior performance compared to SA and SS designs, albeit at the expense of an increased number of RF chains. In contrast, SA and SS configurations enable reduced-complexity PASS deployments. This yields that hybrid schemes combining these approaches could offer a balanced solution, providing near-optimal performance while mitigating hardware and computational complexity.  

\item The adoption of an in-waveguide attenuation model and probabilistic LoS results in an accurate and realistic evaluation of the achievable rate in PASS. Essentially, the former assumption can yield limitations at higher frequency bands, where the in-waveguide attenuation can be more pronounced.

\end{itemize}

\section{Pinching Antenna Systems for Power-Efficient Communication Systems}
\label{passpoweff}

\subsection{PASS for Power Minimization}

In \cite{passpowmin1}, the authors presented an analytical framework for the electromagnetic propagation profile of PASS. In addition, from a communication perspective, a downlink PASS was analyzed, featuring multiple waveguides and several served users. A proposed scheme was proposed for optimally positioning the set of PAs and configuring the beamforming vectors, with the objective of minimizing the total transmission power.
The authors of \cite{passpowmin3} proposed a power minimization scheme for a single-waveguide, single-PA, and multi-user downlink PASS. In this work, an uncertain user location model is considered, which assumes the potential presence of each user with a circular area of a given radius. Thus, the optimization framework aims at allocating adequate power levels to each user, guaranteeing, at least, a threshold outage probability, and minimizing the total power, under a user location uncertainty. A more generalizing setup was analyzed in \cite{passpowmin4} by considering multiple PAs pinched along a transmit waveguide. The proposed scheme was applied to an orthogonal multiple access (OMA) scheme. Also, to simplify the PA positions allocation, the original power minimization problem was detailed into three subproblems solved iteratively, namely (i) deriving an optimal number of PAs to activate per each served user, (ii) optimizing the location of the already-derived number of PAs, and (iii) determine each PAs radiation power. While simplifying the analysis and design by considering an equal-power model among the set of PAs, such a scheme requires a continuous repositioning of the set of PAs when the next user is scheduled for being served. The authors of \cite{passpowmin5} tackled the analysis of multi-waveguide PASS serving various DL users. In this work, one of the authors incorporated a practical challenge in PASS: \textit{The probabilistic presence of a LoS blockage}. Such an impairment is likely to happen in dynamic networks with mobile users and/or scatterers. Similar to \cite{passpowmin4}, a time-division multiple-access (TDMA)-based OMA scheme was considered to serve the different users. The probabilistic presence of a LoS between each PA and each user is modeled by a Bernoulli-distributed binary random variable embedded in the complex-valued channel coefficient of the respective link. Thus, an optimization framework was proposed to optimize the PA locations and transmit beamforming vector minimizing the total power, subject to a minimal average SNR constraint. 

While the aforementioned research work have responded to the question of minimizing the consumed power subject to certain rate constraints, they overlooked a crucial element in terms of the overall power consumed: \textit{The power consumption due to the mechanical or electrical PA movement}. Such an additional component creates an inherent trade-off between moving PAs to favorable positions in terms of communication rate and the amount of power required to position a PA, particularly when favorable locations require more power to move the PAs. In \cite{passpowmin2}, a multi-waveguide multi-user downlink PASS was studied. In particular, an optimization scheme was proposed to minimize the total power consumption, formulated as the sum of (i) the transmit signal power and (ii) the PA displacement one. In addition, the proposed scheme optimize the portion of power radiated by each PA from the total power available at the waveguide. By utilizing an alternating direction method of multipliers (ADMM), a BCD scheme, and an SCA technique, the transmit beamforming vectors, PA radiation power portion, and PAs' positions were optimized.

The work reviewed above has tackled the challenging optimization problems, often involving the joint optimization of the transmit beamforming vectors as well as the PA locations. Thus, techniques such as AO, ADMM, and SCA offer a viable means to solve the original problem by virtue of a tractable substitute convex problem, while the various variables are solved alternately. Nonetheless, such techniques usually offer local optimal solutions, whereby the gaps with respect to the global optimal one is unidentified. To fill this gap, the authors in \cite{passpowmin6} analyzed multi-waveguide multi-user downlink PASS. The analyzed setup considers a discrete PA activation, i.e., the set of PAs can be positioned only on a preset number of positions. For fixed positions of the set of antennas, a power minimization problem was formulated, aiming at finding the optimal binary activation status of each PA as well as the corresponding beamforming matrix. For the single-user scenario, a closed-form expression for the beamforming vector was adopted, corresponding to the maximal-ratio transmission (MRT) one, while a branch-and-bound (BnB)-based algorithm was adopted for identifying the binary PA activation matrix. For the multi-user scenario, an adapted BnB-based approach is proposed, considering the set formed by the activation and beamforming matrices as a basis for constructing the BnB bounding boxes. Notably, the obtained solution converges to a global optimal one, however, after thousands of iterations. The work in \cite{passpowmin7} investigates  PASS serving an uplink unmanned aerial vehicle (UAV)-based network. The analyzed network assumed several UAV user nodes connected to a BS to offload their computational tasks. The BS is extended through a waveguide with several receive PAs, and is linked to a mobile edge computing server (MEC) to execute the computation tasks. Thus, the total power in this system involves the UAV offloading (uplink transmission) power, the UAV propulsion power, and the MEC server local computation power. An optimization problem is formulated, seeking the minimization of the aggregate power consumed by optimally tuning the PA locations, the UAVs' transmit power levels, the UAV trajectory, and the offloading fraction.

\subsection{PASS for Energy-Efficiency Maximization}

In \cite{passeneff1}, Qian \textit{et al.} provided a thorough analysis of the energy efficiency (EE) and the spectral efficiency (SE) of a downlink PASS. Notably, two deployment topologies were analyzed: a centralized deployment in which the PAs are pinched into one of the waveguides to serve users in a TDMA fashion, and a distributed deployment in which a single PA is assigned to each waveguide for simultaneous multiuser transmission. On the one hand, the results unveiled the superiority of the centralized deployment vs. the distributed one in terms of EE, due to the involvement of several RF chains in the distributed one. On the other hand, the distributed deployment results in a notable SE gain compared to the centralized one, due to its inherent achievable multiplexing gain. The authors of \cite{passeneff2} aimed at analyzing the same SE-EE trade-off in PA-aided networks, essentially single- and multi-user deployments. By constructing a novel metric encompassing both the SE and the EE, utilized as an objective function to maximize. Therefore, two distinct optimization problems are expressed for the single- and multi-user scenarios. While the former scenario benefits from a closed-form solution for the transmit beamforming and PA locations, the latter requires an AO- and PSO-based approach to optimize the aforementioned variables. The results show that the SE-EE trade-off gain widens with respect to a traditional MIMO system when increasing either the coverage area or the number of users. The work in \cite{passeneff3} introduces four different deployment PASS schemes for a multi-user single-waveguide DL network. The proposed four PA deployment approaches are established using coarse PA positioning by means of either sliding or selecting a base containing one or many PAs. Such a coarse base location optimization is followed by either a PA activation or location fine-tuning, producing the four proposed topologies. Such architectures provide a wider spectrum of varying conflicting metrics, namely, the SE, power consumption, and the response time needed to reconfigure the PASS. A novel power consumption model was developed involving the power consumed by the motorized actuator, the piezoelectric motor, and the PA activator to assess the EE of each of the four topologies. Results identified the Base selection and PA tuning (SAT) scheme as the most robust in terms of EE.  

A single-waveguide DL PASS was analyzed in \cite{passeneff4}, serving multiple users by means of a TDMA multiplexing scheme. The proposed scheme aims at maximizing the system's EE based optimizing the power allocated to each user, taking into account a per-user minimal rate and PA location spacing constraints.

The potential of AI tools, particularly ML and DL tools, facilitated a robust optimization of diverse wireless systems with complicated, e.g., non-convex, NP-hard optimization problems. To this end, the work in \cite{passeneff5} introduced DL algorithms in maximizing a downlink PASS's EE. The analyzed network consisted of a single waveguide with various PAs pinched, serving multiple users in a TDMA way. A graph attention network-based model is proposed for maximizing the network's EE subject to a power budget and PA location restriction constraints. The adopted graph neural network (GNN) architecture produced optimal PA locations and power levels for the various users' signals. The proposed bipartite GAT demonstrated notable gains against benchmark DL and convex solver-based frameworks.

PASS and RIS share a common potential of reconfiguring the wireless channel and adapting it to the QoS requirements. While their operation principle are different, a comparison between their achievable SE and EE gains is worth exploring. In \cite{passeneff6}, the authors carried out a thorough analytical evaluation of the achievable EE by a RIS-based system and a PASS in a network of two users communicating with each other, under a blocked direct link. The analysis incorporated, in additional to the FSPL, a loss within the waveguide and at the PA while radiating, while the RIS was studied under phase noise impairments. The key observation made is that an RIS requires a huge number of reflective elements (in the order of $10^4$) to match the EE level of a PASS. Such a result is essentially due to the PA contribution in reducing the FSPL, unlike the RIS, which can be placed close to either of the communication ends.

Another TDMA-based EE-efficient PASS scheme was proposed and studied in \cite{passeneff7}, considering a single waveguide. Herein, the authors linked the power radiation of each PA to (i) the coupling between it and adjacent PAs and to (ii) its location on the waveguide. Then, the radiated power model takes into account the remaining power portion within the waveguide. Two EE-maximizing schemes are deployed, namely a static and a dynamic one. On the one hand, the latter scheme finds an optimal solution for the PAs' locations and waveguide feed power, maximizing the instantaneous EE, while the former considers optimal PA locations maximizing the ergodic EE. For both schemes, the coupling length of each PA is maximized also over the ergodic EE. Results showed that the dynamic architecture significantly outperforms the static one, at the cost of a continuous PA locations fine-tuning over for each channel realization (e.g., for every TDMA time slot or whenever a change in the location/channel of a user takes place).

Table \ref{tableeneff} summarizes the work surveyed above on PASS-aided energy-efficient transmission.

\begin{table*}[t]
\centering
\caption{Summary of PASS works on power consumption and energy efficiency.}
\begin{tabular}{|p{2.5cm}|p{2.5cm}|p{7.5cm}|p{3.5cm}|}
\hline
\textbf{Subategory} & \textbf{Work} & \textbf{Contribution} & \textbf{Optimization Technique} \\ \hline

\multirow{8}{*}{Power Minimization}

& Wang et al. \cite{passpowmin1} 
& Studied a downlink multi-waveguide PASS serving multiple users and formulated a transmit power minimization problem. A joint optimization of PA placement and beamforming vectors is conducted. 
& AO, penalty-based optimization, second-order cone, and line-search  \\ \cline{2-4}

& Zeng et al. \cite{passpowmin3} 
& Investigated a single-waveguide single-PA multi-user PASS under user location uncertainty. The framework allocates transmit power ensuring a minimum outage probability while minimizing total power. 
& Bisection method (single-user) and particle swarm optimization (multi-user case)\\ \cline{2-4}

& Li et al. \cite{passpowmin4} 
& Considered multiple PAs along a waveguide for OMA transmission. The total transmit power is minimized via iterative optimization of the number of active PAs, their locations, and their radiation power. 
& Iterative decomposition and closed-form-based optimization \\ \cline{2-4}

& Li et al. \cite{passpowmin5} 
& Investigated multi-waveguide PASS under probabilistic LoS blockage. A joint PA placement and beamforming framework minimizes total power subject to an average SNR constraint. 
& Joint optimization using closed-form solutions and a projected gradient descent. \\ \cline{2-4}

& Xu et al. \cite{passpowmin2} 
& Proposed a comprehensive power consumption model including transmit power and PA movement power. The scheme jointly optimizes PA positions, radiation power portions, and beamforming vectors. 
& ADMM, BCD, SCA \\ \cline{2-4}

& Xu et al. \cite{passpowmin6} 
& Considered discrete PA activation with fixed candidate locations. A power minimization problem was solved by optimizing binary PA activation and beamforming matrices, achieving global optimality. 
& BnB\\ \cline{2-4}

& Ai et al. \cite{passpowmin7} 
& Studied uplink PASS-assisted UAV networks with MEC task offloading. The total system power includes UAV transmission, propulsion, and computation power, jointly optimized with PA placement and UAV trajectory. 
& Block coordinate descent, linear programming. \\ \hline

\multirow{14}{*}{\parbox{2.8cm}{Energy Efficiency \\ Maximization}}

& Qian et al. \cite{passeneff1} 
& Analyzed EE and SE trade-offs in DL PASS for centralized and distributed PA deployments. Results show higher EE for centralized deployment and higher SE for distributed architecture. 
& Analytical evaluation \\ \cline{2-4}

& Zhou et al. \cite{passeneff2} 
& Proposed a joint SE–EE metric for PASS networks and optimized beamforming and PA locations for both single-user and multi-user scenarios. 
& AO and PSO. \\ \cline{2-4}

& Gan et al. \cite{passeneff3} 
& Proposed four PA deployment architectures and developed a detailed power consumption model including actuator and activation power to evaluate EE. 
& Heuristic deployment optimization \\ \cline{2-4}

& Zeng et al. \cite{passeneff4} 
& Investigated a TDMA-based multi-user PASS aiming to maximize EE by optimizing user power allocation under minimum rate and PA spacing constraints. 
& Convex optimization (power allocation) \\ \cline{2-4}

& Xie et al. \cite{passeneff5} 
& Proposed a DL framework using a graph attention network to maximize PASS EE by jointly learning PA locations and user power allocation. 
& Graph Attention Network (deep learning) \\ \cline{2-4}

& Samy et al. \cite{passeneff6} 
& Compared the achievable EE of PASS and RIS-assisted systems under blocked direct links. The study highlights that RIS requires a very large number of elements to match PASS EE performance. 
& Analytical comparison \\ \cline{2-4}

& Asaad et al. \cite{passeneff7} 
& Proposed static and dynamic EE-maximizing PASS schemes considering PA coupling and waveguide power distribution. Dynamic schemes optimize PA locations and feed power for instantaneous EE. 
& Water-filling and a search-based method.\\ \hline
\end{tabular}
\label{tableeneff}
\end{table*}

\subsection{Lessons Learned}

\begin{itemize}
    \item While the schemes above dealt with EE or power minimization regarding the transmit power or PA movement power, they overlooked the execution power induced by running the proposed algorithms. Essentially, some techniques, such a BnB and AO, can yield global optimal solutions when involved, however, at a considerable computational cost.
    \item While distributed PASS deployments result in a considerable SE boost due to their inherent spatial multiplexing capabilities, the resulting EE is notably higher due to the use of various RF chains.
    \item While findings in \cite{passeneff2} demonstrated the efficacy of PASS against traditional 
    MIMO schemes in larger coverage areas, they present an overlooked trade-off with respect to the in-waveguide loss, as shown in \cite{waveguideloss}. Such a loss creates an additional trade-off between coverage area and EE, which may be more pronounced at higher frequency bands, i.e., highly lossy waveguides at these bands. 
    \item The adoption of a higher number of waveguides offers spatial multiplexing capabilities, which comes at the cost of an increased power consumption by activating multiple RF chains. Thus, a hybrid scheme design can balance the implementation cost by activating only a limited number of PAs and RIS reflective elements to ensure a better EE.
    \item As RIS and PASS share remarkable gains in reshaping the wireless channel, their interplay is worth investigating and. For instance, a futuristic wireless network can feature PASS and RIS to adaptively circumvent the channel's LoS absence while benefitting from the RIS beamforming gain whenever possible.
    
\end{itemize}

\section{Pinching Antenna Systems for Secure Communications}
\label{passpls}
Security-wise, traditional collocated antenna arrays at the base station can generate spatially separable signal beams to serve various legitimate users, while maintaining these signals out of the reach and/or the decoding capability of malicious eavesdroppers. Nonetheless, the

\subsection{PLS For Confidentiality}

The authors of \cite{passpls1} introduced the concept of PASS for establishing a secure transmission. Precisely, the considered setup consists of a single waveguide and a single PA. The latter can be flexibly positioned along the waveguide. The work provided an analytical expression for the system's secrecy outage probability (SOP) as a function of the PA, user, and eavesdropper's coordinates. Despite the initial insights obtained, the analyzed network was limited to a simplistic setup, where the analysis does not provide a solution for the optimal PA position. The authors of \cite{passpls3} proposed a secure single-waveguide-based PASS scheme considering a discrete PA position-based setup. By positioning the $N$ PAs uniformly along the waveguide, a coalitional game-based antenna activation algorithm is developed to activate only a subset of PAs that cooperatively maximize the system's secrecy capacity (SC).  Furthermore, in \cite{passpls4}, Zhong \textit{et al.} proposed a single-waveguide secure PASS utilizing index modulation (IM) and directional modulation (DM). The proposed scheme considers multiple cooperating legitimate users and collaborative eavesdroppers, where a single legitimate user is scheduled at each transmission time slot. IM is carried out by conveying part of the binary sequence by means of the corresponding scheduled legitimate user, while artificial noise (AN) and DM are utilized to deliberately distort the constellation at the malicious eavesdroppers. Then, based on the designed AN signal, the optimal PAs locations are determined to be the nearest to the scheduled legitimate user, ensuring a constructive phase alignment of radiated signals from the set of PAs.

In \cite{illipinching}, a secure PA-aided scheme is proposed and evaluated. The considered scheme represents a generalization of the one in \cite{passpls1} by considering a multi-waveguide network with multiple PAs pinched on each one. By virtue of pinching beamforming, legitimate signal beams can be designed to preserve preset secrecy requirements of the system defined by a minimal legitimate SINR and a maximal eavesdropping SINR constraints. Thus, an optimization problem to minimize the total transmit power, subject to the secrecy constraints, is formulated and solved by virtue of semidefinite programming (SDP) and semidefinite relaxation (SDR) to obtain the optimal beamforming vectors for a given setup of PAs equidistantly spaced.

In \cite{passpls2}, a secure multi-waveguide PA-aided scheme is proposed to guarantee data confidentiality. The proposed scheme extends the setups in \cite{passpls1,passpls3} by considering multiple waveguides. Also, the scheme in \cite{passpls2} is investigated under two scenarios, namely a (i) single user and eavesdropper or (ii) multiple ones. For the former setup, a closed-form solution for the optimal beamforming vector maximizing the system's SC is provided, while an element-wise BCD-based approach was proposed for optimizing the PAs' positions over the different waveguides. On the other hand, an alternating optimization (AO)-based framework leveraging fractional programming (FP), Lagrange multipliers, and Gauss-Seidel techniques is developed for optimizing the pinching beamforming vectors and PA locations sequentially to maximize the sum SC (SSC). The authors in \cite{passpls5} proposed two secure PA-aided schemes employing a single and two waveguides. The former scheme proposes a novel PA positioning algorithm, labeled as PA-wise successive
tuning (PAST), which aims at forcing a constructive signal supersposition at the legitimate receiver and a destructive one at the eavesdropper. This is established by controlling the per-PA radiated signal phase, depending on the PA position. To this end, PAST is based on an initial coarse PA positioning, minimizing the FSPL with respect to the legitimate user. Then, it optimizes the PA locations starting from PA in the middle, until completing the set of PAs. On the other hand, the multi-waveguide scheme broadcasts both an AN signal and the confidential one to enhance the system's secrecy. Two strategies are proposed, namely (i) waveguide division (WD), based on dedicating a waveguide to each type of signals, and (ii) waveguide multiplexing (WM), multiplexing both types of signals on each waveguide. While PAST can be utilized for WM, an AO-based framework is developed for the WM strategy by relying on the particle-swarm optimization (PSO) heuristic-based algorithm for PAs optimal positioning, while SCA was used for retrieving the optimal beamforming and AN covariance matrices.

The authors of \cite{passpls7} investigate the secrecy performance of a single-PA single-waveguide PASS in the presence of an eavesdropper. An analytical approximation for the secrecy outage probability (SOP) is derived, considering a random locations of both the legitimate user and the eavesdropper. The asymptotic analysis of the SOP is further analyzed, revealing the existence of a lower bound for the PASS, which is significantly smaller than that of conventional fixed-position antenna systems. Another similar secure PASS design was investigated in \cite{passpls6}. However, the considered setup consists of various PAs placed at fixed positions, from which only a subset is activated. The considered setup extends the definition of eavesdropping assumptions into an upper-layer statistical security, i.e., data is likely deciphered by the eavesdropper if the decoding queue length exceeds a given threshold. Thus, an optimization problem is formulated to maximize the legitimate transmission rate, subject to a total power and effective eavesdropping bandwidth. An optimal solution for the power allocation level and PA activation pattern is proposed.

The proposed optimization techniques in the aforementioned work provided acceptable solutions of the underlying optimization problems, balancing security, reliability, and power consumption requirements. However, one should note that the presence of secrecy constraints renders the convergence of iterative model-based methods' either slow or not guaranteed. On the other hand, heuristic-based method (e.g., PSO) are very sensitive to the initial guess. Thus, as an effort to improve PASS' secrecy performance, the authors of \cite{passpls8} adopted a GML to minimize the total transmit power, subject to a maximal outage probability (OP) and a minimal secrecy rate (SR) constraints. By considering a single PA, single waveguide, and user location uncertainty, the proposed GML-based approach learns the gradient of the loss function composed by the legitimate rate and SR. As a result, GML stands out as an adaptive solution for network parameter changes, while outperforming conventional model-based optimization frameworks.

\subsection{PLS For Covert Communications}

Despite the growing interest in PASS deployment for realizing confidential transmission, only two works were reported so far on covert communication, i.e., \cite{passcovert1,passcovert2}. In \cite{passcovert1}, the authors studied a covert PASS employing either a single waveguide with a single PA (SWSP), or multiple waveguides with multiple PAs (MAMP). Such a covert PASS offers additional degrees of freedom in realizing a covert transmission by virtue of the flexible PA positioning. Importantly, the studied model considers an uncertainty error on the Warden node's position, originating from its passive nature and the limited information about it. The authors introduced the notion of forbidden zone, in which a covert transmission cannot be ensure. Thus, for the SWSP architecture, a closed-form expression for the optimal PA position was provided, which enables obtaining an iterative-based solution for the optimal transmit power. For the MAMP scheme, a twin particle-swarm optimization (PSO) framework is designed to optimize the initial positions to determine an optimal beamforming as well as the PA locations.

While the abovementioned work provided several insights on PASS potential for realizing a covert transmission, its main issue from a practical perspective lies in obtaining an accurate channel state information about the warden nodes. Thus, in \cite{passcovert2}, the authors proposed a sensing-aided covert transmission scheme. In this work, the authors proposed a sensing-powered estimation of the adversary's position, which enables constructing its channel state information (CSI). In addition, due to the mobility of the wardens, an extended Kalman filter (EKF)-based mechanism is employed for an accurate real-time tracking of the wardens. The authors formulated an optimization problem maximizing the covert transmission rate, subject to covertness and sensing constraints. An iterative algorithm is proposed, decomposing the original problems into subproblems optimizing the legitimate signal beamforming vectors, the artificial noise process, and the PA positions. Precisely, while a closed-form solution was provided for the beamforming and AN vectors, a deep reinforcement learning (DRL) scheme was employed to optimize the PA positions, adopting a soft actor-critic approach. The authors of \cite{passcovert3} studied a covert backscattering communication-based PASS. A pair of transmit and receive waveguides are considered, equipped with a single PA each, which are used for transmitting the signal and collecting the scattered one from a backscattering device. Closed-form solutions for the optimal transmit power and the transmit PA location, maximizing the covert transmission rate, are provided.

Table \ref{tablepls} provides a concise view of the above-reviewed work for secure PASS, covering both confidentiality and covert communication aspects.

\begin{table*}[t]
\centering
\caption{Summary of PASS works for PLS.}
\label{tablepls}

\begin{tabular}{|p{2cm}|p{2.5cm}|p{9cm}|p{3cm}|}
\hline
\textbf{Category} & \textbf{Work} & \textbf{Contribution} & \textbf{Optimization Technique} \\ \hline

\multirow{8}{*}{Confidentiality}

& Badarneh et al. \cite{passpls1}
& Introduced PASS for secure communication using a single waveguide and a single movable PA. An analytical expression for the SOP was derived as a function of the PA, user, and eavesdropper locations.
& Analytical evaluation \\ \cline{2-4}

& Wang et al. \cite{passpls3}
& Proposed a secure single-waveguide PASS with uniformly spaced PAs. A subset of antennas is activated to maximize secrecy capacity through cooperative transmission.
& Coalitional game-based antenna activation \\ \cline{2-4}

& Zhong et al. \cite{passpls4}
& Developed a secure PASS scheme combining IM, DM, and artificial noise to distort eavesdroppers’ constellation while enhancing the legitimate signal.
& PA location optimization based on AN design \\ \cline{2-4}

& Illi et al. \cite{illipinching}
& Proposed a multi-waveguide PASS using pinching beamforming to satisfy secrecy SINR constraints at legitimate users while limiting eavesdroppers’ SINR. The setup considered fixed PA locations.
& SDP and SDR for beamforming optimization \\ \cline{2-4}

& Sun et al. \cite{passpls2}
& Studied secure multi-waveguide PASS for single- and multi-user scenarios, maximizing the SC or sum sum SC through joint beamforming and PA positioning.
& BCD, AO, and Fractional Programming \\ \cline{2-4}

& Zhu et al. \cite{passpls5}
& Proposed secure PASS architectures with single and multi-waveguide setups. Introduced the PAST algorithm and artificial noise transmission strategies.
& PAST (heuristic-based), PSO, and SCA. \\ \cline{2-4}

& Li et al. \cite{passpls7}
& Investigated secrecy outage performance of a single-PA PASS under random user and eavesdropper locations and derived asymptotic SOP bounds.
& Analytical evaluation \\ \cline{2-4}

& Wang et al. \cite{passpls6}
& Considered fixed PA positions with selective activation and formulated an optimization maximizing legitimate rate under statistical eavesdropping constraints.
& Closed-form analysis and a matching theory-based algorithm\\ \hline

\multirow{3}{*}{Covertness}

& Jiang et al. \cite{passcovert1}
& Studied covert PASS with either single-waveguide single-PA or multi-waveguide multi-PA architectures under warden location uncertainty. Introduced the concept of forbidden zones for covert transmission.
& Closed-form optimization and PSO \\ \cline{2-4}

& Jiang et al. \cite{passcovert2}
& Proposed sensing-assisted covert PASS where the adversary position is estimated and tracked using sensing and EKF. The covert rate is maximized subject to covertness constraints.
& Iterative optimization and DRL (Soft Actor-Critic) \\ \cline{2-4}

& Wang et al. \cite{passcovert3}
& Investigated covert backscatter PASS with transmit and receive waveguides. Closed-form solutions for optimal PA placement and transmit power maximizing covert rate were derived.
& Closed-form optimization \\ \hline

\end{tabular}
\end{table*}

\subsection{Takeaways}

The above surveyed secure PASS schemes can be decomposed into two distinct categories, namely single waveguide-based architectures and multi-waveguide-based ones. On one hand, single-waveguide-based strategies are usually based on phase alignment of the signal radiated from the different PAs to either create a constructive signal superposition at genuine receivers or a destructive one at Eves. While such a scheme results in a lower complexity by requiring only a single RF chain, its performance can be limited in some critical cases. For instance, smart (active) eavesdroppers can adapt their position along the $x$ and $y$ axes, as can be inferred from \cite[Eqs. (15)-(16)]{passpls5} to force similar received superposed signals at both the legitimate user and the eavesdropper. On the other hand, multi-waveguide approaches can mitigate such an observed limitation in single-waveguide setups by increasing the propagation degrees of freedom through pinching beamforming. This allows for directing a

\section{Pinching Antenna Systems for Multiple Access}
\label{passmultacc}
Multiple access schemes can be categorized generally into orthogonal and non-orthogonal ones. The former categories of schemes allocate a dedicated resource unit (e.g., Frequency subcarrier, time slot, spreading code) to each user to avoid inter-user interference, while the latter aims at establishing a multi-user communication utilizing the same resource. In such techniques, the overarching idea is to design efficient methods to mitigate interference, thereby preserving acceptable communication rates.

Non-orthogonal multiple access (NOMA) and rate-splitting multiple access (RSMA) represent the two key techniques in the categories of non-orthogonal techniques. Over the past decade, considerable research efforts have been made to investigate and optimize the performance of NOMA and RSMA techniques over different network setups, mainly over single-input single-output (SISO) schemes as well as MIMO ones, considering traditional uniform linear arrays (ULAs). 

From a PASS perspective, the consideration of a PA-enabled flexible antenna design can help in reshaping the propagation channel per the need. 

\subsection{Related Work Review}

\subsubsection{NOMA}

The seminal work of Wang \textit{et al.} in \cite{passnoma1} introduced the PASS concept in a downlink NOMA network, where multiple antennas are pinched on a single waveguide. The developed scheme focused on finding the optimal number of activated PAs along with their locations in order to maximize the sum rate. By developing a low complexity matching theory-aided algorithm, the sum rate was maximized, and the results indicate an increase in performance compared to a distance-based antenna activation benchmark scheme. In \cite{passnoma2}, an extension to the previous work was conducted by essentially expanding the setup to a multi-waveguide one. Such a setup results in an enhanced system performance by increasing
spatial coverage, which consequently reduces the distances between users and the serving PAs, i.e., creating favorable channel conditions. The proposed PA-NOMA design is based on associating each waveguide to one or more users. Thus, such an approach implies finding an optimal power allocation strategy and user-waveguide pairing to maintain the successive interference cancellation (SIC) procedure unaffected and maximize the sum rate of the network. A coalitional game-based algorithm was developed for optimizing the waveguide-user association, while a polyblock outer approximation algorithm was proposed for optimal power allocation. 
In \cite{passnoma3}, an uplink PA-aided NOMA scheme was studied, consisting of a single receiving PA pinched on a single waveguide. With the aim of maximizing the users' uplink sum rate, two optimization frameworks were developed for two cases, namely when considering a per-user minimal quality-of-service (QoS) and when relaxing it. The latter scheme was tackled by using a particle-swarm optimization (PSO) algorithm for finding the optimal PA locations when the maximal transmit power is used by each user. On the other hand, the former QoS-constrained scheme is solved utilizing an alternating optimization (AO)-based scheme relying on a closed-form solution for the optimal power allocation. Likewise, the authors in \cite{passnoma4} analyzed a similar downlink PA-NOMA setup employing a single waveguide. Precisely, the Karush-Kuhn-Tucker (KKT) are employed on the constructed Lagrangian function to determine the optimal power allocation coefficients among the users' signals, while an iterative two-stage optimization framework was proposed for optimizing the PAs positions. In \cite{passnoma11}, an enhancement on the downlink PA-NOMA scheme of \cite{passnoma4} was proposed and studied, where multiple waveguides were considered for serving several users. The architecture assumes a single PA to activate on each waveguide, which, consequently, enables the possibility of pinching beamforming to direct signals into the subspace spanned by each user's channel. Importantly, the authors introduced in this work an in-waveguide attenuation model to account for the propagation loss between the feed point and the PA, which accentuates at higher frequency bands. Thus, such an impairments creates an inherent trade-off between network coverage and signal power loss (performance degradation). To maximize the sum rate of the network, an optimization problem was formulated for optimizing the beamforming vectors and PA locations, which was decomposed into two subproblems, solved efficiently by virtue of a semidefinite relaxation (beamforming), and an SCA-based approach, along with the use of auxiliary variables for optimal PA positioning. In another scheme \cite{passnoma6}, instead of maximizing the sum rate, the proposed scheme focused on maximizing the decoding rate of a given user while fulfilling a minimal rate constraint for the other in a two-user NOMA setup. A BCD scheme was utilized to decompose the non-convex problem due to the coupling between the power coefficients and the PA locations-dependent channel vectors. Thus, while an optimal solution for the power allocation is obtained, an SCA-based approach alongside the use of auxiliary variables is utilized to convexify the PA locations subproblem. In particular, a closed-form expression for the optimal PA position is given for the single-PA particular case. 

The complex optimization problems in the above-discussed work renders the obtained solution suboptimal ones due to the use of surrogate expressions for the objective or constraint functions. Motivated by this limitation, the work in \cite{passnoma13} proposed a DL-based optimization framework ofa multi-user PASS-NOMA. In particular, a convolutional neural network (CNN) model was trained to learn the optimal set of PA locations maximizing the minimal per-user rate.

Zeng \textit{et al.} proposed in \cite{passnoma5} an energy-efficient uplink PA-NOMA scheme. Unlike the abovementioned works, which aimed at maximizing the network sum rate, the current work tackled EE maximization. The scheme consists of a single waveguide with a single receive PA. An AO-based algorithm was proposed for finding an optimal transmit power per user and the PA position, maximizing the network's EE. While the former subproblem (power allocation) was solved by the Dinkelbach algorithm, a PSO-based scheme was developed for optimally positioning the PA. Similarly, another work in \cite{passnoma8} investigated the power minimization problem in a PA-NOMA network utilizing multiple waveguides. For a number of users equal to the number of PAs per waveguide, the scheme considers assigning a user's signal to each PA for radiation. For an equidistant positioning of PAs, a framework was developed to optimize the power allocation factor per antenna and the power budget per waveguide, with the objective of minimizing the overall system power. To this end, a closed-form expression for the optimal power allocation strategy was presented, indicating a power decrease by one order of magnitude compared to an equal-power baseline scheme. Likewise, a similar scheme was recently proposed in \cite{passnoma10}, aiming also to minimize the total power, however, in a downlink PA-NOMA system. By focusing on a single-waveguide setup, the proposed scheme targeted network power minimization subject to a minimal decoding rate per user. While the problem has been proven to be convex and readily solvable by any convex optimization solver, two closed-form expressions are given for an efficient evaluation, exploiting the structure of the rate constraints in the tackled problem. In \cite{passnoma7}, the authors considered a multi-waveguide downlink PA-NOMA setup, with the objective of minimizing the total system power. The network considers a cluster-based design where users are grouped to create favorable channel conditions in each cluster. Thus, a user grouping algorithm is proposed based on channel and location correlations. Then, a power minimization problem subject to a minimal per-user rate constraint is formulated. Thus, two solution procedures are proposed, namely (i) a gradient-based on utilizing a majorization minimization (MM) and penalty dual decomposition (PDD) methods to tackle the non-convex problem, and (ii) a heuristic PSO-based scheme for optimal PA positions combined with a suboptimal zero-forcing beamformer.

Partial NOMA (P-NOMA) is a variant of NOMA that allows the various users to simultaneously use only a portion of the subcarriers, while the other portion is shared to avoid interference, as done in orthogonal multiple access (OMA). Thus, the overlap interval between the served users can be controlled to balance the decoding performance. Thus, the authors of \cite{passnoma9} proposed a PA-aided P-NOMA design, aiming at leveraging the aforementioned benefits of P-NOMA, in addition to the channel tunability offered by PAs, in order to maximize the sum rate of a two-user downlink network. The analysis provides a closed-form expression for the optimal PA placement in a PA-based P-NOMA setup, consisting of a single PA, while the achievable sum rate is analyzed and compared with PA-OMA and PA-NOMA schemes. The results demonstrate the great potential of PAs and PA NOMA to boost the sum rate, compared to classical NOMA and OMA schemes. In \cite{passnoma12}, the sum rate of a single-waveguide PA-NOMA network was maximized. The setup consists of several equidistant PAs placed along the waveguide. By virtue of an AO-based framework, the optimal NOMA signals power fraction is derived, while an SCA-based procedure is proposed to optimize the radiated power by each PA.

\subsubsection{RSMA}

It is worth mentioning that the related literature in PA-aided RSMA networks, i.e., PA-RSMA, is limited compared to its NOMA counterpart. Up to date, only four related work that are present, namely \cite{passrsma1,passrsma2,passrsma3,passrsma4}. 

The work in \cite{passrsma1} considered a two-user uplink PA-RSMA system, where a single waveguide is equipped with two PAs, each serving a user in separate environments. A closed-form expression for the system's outage probability (OP) was derived, demonstrating the notable gains of PA-RSMA against PA-NOMA and PA-OMA schemes. Despite missing an optimization aspect of the operating PAs in the system, such an analysis provided notable insights on the channel capacity region of PA-RSMA systems.

The authors of \cite{passrsma2} considered a downlink PA-RSMA system consisting of several waveguides with multiple transmit PAs activated on each. A proposed method for PAs placement was proposed, based on sequentially positioning the set of PAs in the system in order to minimize the inter-user channel correlation and PAs-users sum distance. Then, based on such a proposed placement strategy, the optimal beamforming vectors are optimized by virtue of some introduced auxiliary variables along with semidefinite relaxation (SDR). The obtained results demonstrate a the superiority of the proposed scheme against a baseline one based on activating all the antennas as well as against an RSMA scheme based on a conventional collocated ULA.

The authors of \cite{passrsma3} analyzed a PA-RSMA system serving various downlink users utilizing a single waveguide and a single PA. The serving BS maitains a content library of various files requested frequently by users. Herein, a content-aware strategy is introduced to limit the number of private streams when multiple users request the same content from the BS. Accordingly, an optimization problem is formulated with the objective of minimizing the average system latency, subject constraints on the total power, the common stream decoding rate, and SIC feasibility. An AO-based framework was proposed to optimize the common stream rate, per-stream power allocation, and PA position. 

In \cite{passrsma4}, a single-waveguide uplink PA-RSMA scheme was studied. The setup consists of multiple received PAs pinched along the waveguide for serving several uplink users. An AO-based framework is designed by decoupling the optimization of both the PA locations and the users' transmit power levels into two distinct subproblems. By virtue of an SCA- and Taylor expansion-based surrogate functions, concave bounds of the per-user rate are obtained, which render the subproblems readily solvable by convex optimization tools.

Table \ref{tablemultacc} provides a summary of the above-reviewerd work on PASS-enabled multiple access schemes.

\begin{table*}[t]
\centering
\caption{Summary of PASS-enabled multiple access schemes.}
\label{tablemultacc}
\begin{tabular}{|p{2cm}|p{2.5cm}|p{9cm}|p{3cm}|}
\hline
\textbf{Category} & \textbf{Work} & \textbf{Contribution} & \textbf{Optimization Technique} \\ \hline

\multirow{12}{*}{NOMA}

& Wang et al. \cite{passnoma1}
& Introduced PASS in a downlink NOMA network with multiple PAs on a single waveguide. The objective was maximizing the network sum rate by selecting the optimal PA subset and their locations.
& Matching-theory-based algorithm \\ \cline{2-4}

& Wang et al. \cite{passnoma2}
& Extended PA-NOMA to a multi-waveguide architecture. The framework jointly optimizes user–waveguide association and transmit power allocation to maximize the system sum rate while ensuring proper SIC decoding.
& Coalitional game + Polyblock outer approximation \\ \cline{2-4}

& Zeng et al. \cite{passnoma3}
& Investigated an uplink PA-NOMA system with a single receiving PA. Two optimization frameworks were proposed for sum-rate maximization with and without QoS constraints.
& PSO and AO \\ \cline{2-4}

& Zhou et al. \cite{passnoma4}
& Studied a downlink PA-NOMA system with a single waveguide. Optimal power allocation was derived using KKT conditions while PA locations were optimized through an iterative two-stage framework.
& KKT-based optimization and iterative search \\ \cline{2-4}

& Hu et al. \cite{passnoma11}
& Proposed a multi-waveguide PA-NOMA scheme incorporating in-waveguide attenuation effects. The beamforming vectors and PA locations were jointly optimized to maximize the sum rate.
& SDR and SCA \\ \cline{2-4}

& Xu et al. \cite{passnoma6}
& Considered a two-user PA-NOMA system maximizing the decoding rate of one user while ensuring a minimum rate for the other.
& BCD and SCA \\ \cline{2-4}

& Zeng et al. \cite{passnoma5}
& Proposed an energy-efficient uplink PA-NOMA scheme optimizing transmit power and PA position to maximize energy efficiency.
& AO, Dinkelbach algorithm (fractional programming), and PSO \\ \cline{2-4}

& Fu et al. \cite{passnoma8}
& Investigated power minimization in a multi-waveguide PA-NOMA system with equidistant PA placement.
& Closed-form power allocation \\ \cline{2-4}

& Mohammadzadeh et al. \cite{passnoma10}
& Studied a downlink PA-NOMA system aiming to minimize total transmit power while satisfying minimum user rate constraints.
& Convex optimization + closed-form solution \\ \cline{2-4}

& Gan et al. \cite{passnoma7}
& Proposed a cluster-based multi-waveguide PA-NOMA architecture. Users are grouped based on channel/location correlation to reduce total system power.
& MM, PDD, PSO \\ \cline{2-4}

& Zhuo et al. \cite{passnoma9}
& Introduced a PA-based partial NOMA (P-NOMA) system where users partially share subcarriers. A closed-form PA placement maximizing the sum rate was derived.
& Closed-form optimization \\ \cline{2-4}

& Vashakidze et al. \cite{passnoma12}
& Maximized the sum rate in a PA-NOMA network with equidistant PAs along a waveguide by optimizing NOMA power fractions and PA radiated power.
& AO and SCA \\ \hline

\multirow{4}{*}{RSMA}

& Tegos et al. \cite{passrsma1}
& Studied an uplink PA-RSMA system with two users and derived a closed-form expression for the outage probability, showing performance gains over PA-NOMA and PA-OMA.
& Analytical evaluation \\ \cline{2-4}

& Wang et al. \cite{passrsma2}
& Proposed a multi-waveguide downlink PA-RSMA scheme with sequential PA placement to reduce channel correlation and optimize beamforming.
& Sequential PA placement and SDP/SDR \\ \cline{2-4}

& Hua et al. \cite{passrsma3}
& Investigated a content-aware PA-RSMA system where the base station adapts the number of private streams based on requested files to minimize system latency.
& AO \\ \cline{2-4}

& Phyo et al. \cite{passrsma4}
& Studied an uplink PA-RSMA network with multiple receive PAs. The system jointly optimizes user transmit power and PA locations to maximize the achievable rate.
& AO, SCA, and Taylor approximation. \\ \hline

\end{tabular}
\end{table*}

% \subsection{Takeaways}
% The adoption of a PASS-aided design can 

\section{Pinching Antenna Systems for ISAC Networks}
\label{passisac}
% The emergence of sensing-based applications calls for the need to develop robust schemes  .  

\subsection{Related Work Review}

As far as the related literature on PA-ISACS goes, the majority of works have mostly tackled the evaluation and optimization of the sensing-reliability trade-off. The seminal work of Ding in \cite{passisac1} aimed at quantifying PASS's gains in terms of sensing compared to a conventional system employing a uniform array. The studied system consists of a multi-waveguide PASS aiming to communicate and localize multiple users simultaneously. For a sensing perspective, the performance was quantified in terms of the per-user Cram\'er-Rao bound (CRB) of the distance to the user's estimate. For an equally spaced PA-based setup, analytical expressions revealed interesting insights on the inter-PA spacing to minimize the CRB, which was found to be proportional to the waveguide height. 

\begin{figure}[h]
\par
\vspace*{-.3cm}\begin{center}
\includegraphics[scale=.43]{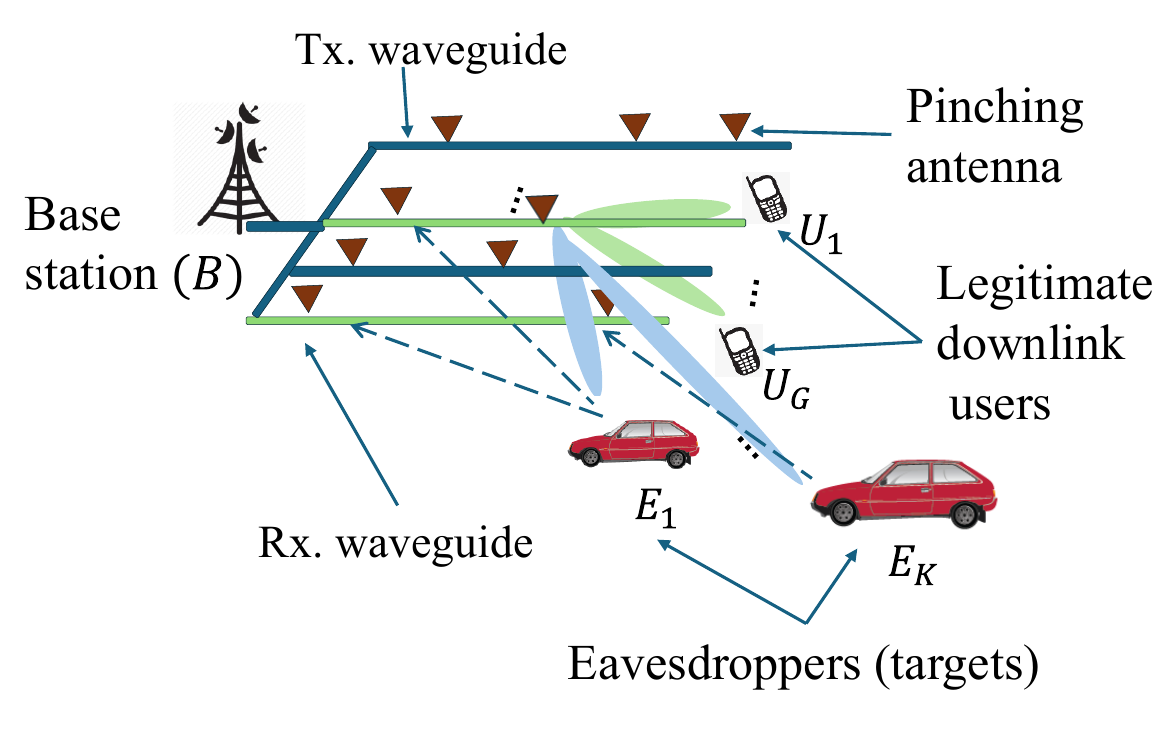}
\end{center}
\par
\captionsetup{font=footnotesize}
\caption{{A pinching-antenna-enabled ISAC system \cite{illipassisacsec}.}}
\label{sysmod}
\end{figure}

Bozanis \textit{et al.} analyzed in \cite{passisac2} a bistatic ISAC network operating with a uniform linear array (ULA) of transmit antennas to illuminate a sensed target, whereas several receive PAs are installed equidistantly on a waveguide to collect separately the reflected echo signal. The paper provides a thorough evaluation of the CRB for both the target range and angle of arrival. The analytical framework considers a realistic near-field condition, which holds in PA-ISAC systems due to the large PA array size. 
Furthermore, the authors in \cite{passisac3} conducted an analysis on a similar PA-ISAC network. Unlike \cite{passisac1,passisac2}, this work aimed at solving one of the critical radar sensing challenges, which is the radar cross-section (RCS)'s dependence on the DFRC base station (BS)'s angular look direction. Thus, the proposed PA-ISAC design is based on multiple PAs placed along a transmit waveguide, while only a single PA is activated at each time slot. The reflected sensing echo signals are received and processed by a separate receive multi-antenna ULA. To assess the sensing performance in the considered network, the authors introduce the outage probability (OP) metric to quantify the probability that the accumulated sensing signal-to-noise ratio (SNR) over multiple time slots falls below a certain threshold. Such a metric manifests some usefulness in the presence of a randomly varying RCS across the different time slots and PAs. To this end, the authors formulate an optimization problem for minimizing the sensing OP subject to a minimal communication rate. To tackle the nonconvex communication and sensing metrics expressions, a Chernoff bound and successive convex approximation techniques were used for providing convex surrogate terms, while the binary PA selection variables are substituted by non-binary variables to simplify the problem solution. A similar single-waveguide ISAC network was studied in \cite{passisac5}, whereby an optimization problem was formulated, aiming at maximizing the sensing illumination power in a given target look direction, subject to a communication rate and total power constraint. A penalty-based alternating optimization (AO) framework was developed to jointly optimize the beamforming vector and the PAs' positions. The beamforming optimization can be represented by a convex semidefinite program (SDP), while the PAs' positions subproblem is tackled by a two-layer penalty-based scheme. An interesting extension was covered in \cite{passisac6}, in which the proposed PA-ISAC was optimized considering multiple downlink users and sensed targets, unlike the aforementioned previous work. Similarly, the proposed scheme aimed at optimizing the power allocation and positions of the various PAs pinched along a transmit waveguide. The authors developed a deep reinforcement learning (DRL) framework for optimizing the aforementioned variables to maximize the communication sum rate subject to a minimal sensing SNR, total power, and inter-PA separation constraints. 

Ouyang \textit{et al.} provided in \cite{passisac8} a thorough analytical investigation of the information-theoretic limits of a PA-ISAC system employing a single waveguide for transmission and another one for reception. By considering a single downlink user served and a single sensed target, multiple PAs were activated along the transmit waveguide, while a single PA was activated in the receive one. The authors conducted an evaluation of the achievable communication rate (CR)-sensing rate (SR) region. Such a set of achievable CR-SR values was determined by optimizing first for the PAs positions using a sequential element-wise one-dimensional search. Such a step determines the inner and outer bands of the CR-SR region of a PA-ISAC network. The obtained results demonstrate that the CR-SR trade-off curve becomes more rectangular, i.e., the possibility of achieving a maximal SR without a compromise in the CR, when both the communication receiver and sensed targets are closer to each other. It is also shown that the proposed PA-based design offers a better communication-sensing trade-off compared to conventional ULA-based and time-sharing-based ISAC network designs. By aiming at maximizing the communication rate subject to a total power, sensing radar, and PA positions constraints, 

%%%%% mutiple waveguides

Unlike single-waveguide-based PA schemes, multiple waveguide-based PA designs offer additional flexibility in signal beamforming and channel reconfiguration. The overarching idea lies in dedicating a radio-frequency (RF) chain to each waveguide, which enables the transmission of different signals from each waveguide. By pinching several antennas on each waveguide, the broadcasted signal power for communication and sensing can be boosted, whereas beamforming along the various waveguides can help in simultaneously directing dedicated communication and sensing signals to the served users and sensed targets. Thus, the scheme in \cite{passisac4} introduced the multi-waveguide-based PA-ISAC scheme, where the considered system consists of several waveguides used to sense a single target and serve a downlink user. The waveguides are divided into transmitting and receiving ones, where, in each waveguide, only a single transmit or receive PA is pinched. By virtue of auxiliary variables introduced, an upper bound for the CR, and the successive convex approximation technique, an optimal solution for the PA positions and the beamforming vector was achieved. In addition, a closed-form solution for the single-PA particular case was formulated. The PA-ISAC in \cite{passisac9} was designed for establishing robust communication and sensing with respect to a low-altitude uncrewed aerial vehicle (UAV). The proposed scheme introduced a PA setup of several transmitting waveguides on the ground, targeting to communication with a moving UAV and dynamically detecting it. Several antennas are pinched on each waveguide, where, potentially, only a subset of them can be activated. Also, a bistatic radar setup is assumed, where a separate radar receiver collects and processes the reflected echoes from the UAV. A robust deep reinforcement learning (DRL) framework is developed based on the proximal policy optimization algorithm, which considers the set of PAs, waveguides, and the UAV as agents interacting with each other to maximize the CR subject to a minimal radar SINR constraint. Recently, Jiang \textit{et al.} analyzed in \cite{passisac7} an ISAC-enabled SWAN PASS. By leveraging the SWAN-PASS architecture introduced in Section \ref{passdatarate}, the segments are isolated, and each segment's feed point is linked with an independent RF chain. The authors analyzed the SWAN scheme in a multi-user single-target ISAC network, where three operating protocols for the SWAN architecture were proposed and studied, namely segment selection, segment aggregation, and segment multiplexing. With an optimal design of the PA locations for each of the three schemes, the obtained results showed a notable gain of SWAN over a conventional PASS scheme.

The secrecy aspect of PA-ISAC systems was tackled uniquely in our work \cite{illipassisacsec}. In this system, illustrated by Fig. \ref{sysmod}, the aim is to securely serve various legitimate downlink users while detecting the presence of multiple malicious targets. The latter node aims at illegally eavesdropping on the legitimate signal. Various transmit and receive waveguides are considered in the analyzed PA-ISAC design with several transmit/receive PAs activated on each waveguide to (i) enhance the radiated ISAC signal power and increase. 

In Table \ref{tableisac}, a summary of the afore-reviewed work on PASS-ISAC is provided.

\begin{table*}[t]
\centering
\caption{Summary of PASS-enabled ISAC works.}
\begin{tabular}{|p{2.5cm}|p{2.5cm}|p{9cm}|p{3cm}|}
\hline
\textbf{Category} & \textbf{Work} & \textbf{Contribution} & \textbf{Optimization Technique} \\ \hline

\multirow{5}{*}{Single-Waveguide}

& Ding et al. \cite{passisac1}
& Introduced PASS for ISAC and analyzed the sensing reliability gain compared to conventional antenna arrays. The sensing performance was quantified using the per-user Cramér–Rao bound (CRB) of distance estimation.
& Analytical evaluation \\ \cline{2-4}

& Bozanis et al. \cite{passisac2}
& Studied a bistatic PA-ISAC network with receive PAs deployed along a waveguide. Analytical CRB expressions for target range and angle estimation were derived under near-field propagation conditions.
& Analytical CRB analysis \\ \cline{2-4}

& Khalili et al. \cite{passisac3}
& Proposed a PA-ISAC system addressing radar cross-section variations. The sensing performance was evaluated via sensing outage probability under varying RCS across time slots.
& Chernoff bound (surrogate function) and SCA optimization \\ \cline{2-4}

& Zhang et al. \cite{passisac5}
& Formulated an optimization maximizing sensing illumination power toward a target direction while ensuring communication rate and power constraints.
& AO, SDP, and penalty-based optimization \\ \cline{2-4}

& Qin et al. \cite{passisac6}
& Extended PA-ISAC to multiple users and targets. The communication sum rate was maximized subject to sensing SNR and PA separation constraints.
& DRL \\ \hline

\multirow{5}{*}{Multi-Waveguide}

& Ouyang et al. \cite{passisac8}
& Investigated the information-theoretic communication-sensing rate (CR-SR) region for PA-ISAC. Results demonstrate improved CR-SR trade-off compared to conventional ULA-based ISAC systems.
& Sequential search optimization \\ \cline{2-4}

& Mao et al. \cite{passisac4}
& Proposed a multi-waveguide PA-ISAC architecture with separate transmit and receive waveguides. An optimization framework jointly determines PA locations and beamforming vectors.
& SCA and convex optimization \\ \cline{2-4}

& Hu et al. \cite{passisac9}
& Studied UAV-assisted PA-ISAC with multiple transmit waveguides and bistatic radar reception. A DRL framework was used to optimize PA activation and beamforming.
& DRL (Proximal Policy Optimization) \\ \cline{2-4}

& Jiang et al. \cite{passisac7}
& Investigated SWAN-based PA-ISAC architectures with segmented waveguides and independent RF chains. Multiple segment operation strategies were evaluated for ISAC performance.
& Analytical KKT-based optimization\\ \cline{2-4}

& Illi et al. \cite{illipassisacsec}
& Proposed a secure PA-ISAC design enabling simultaneous sensing of malicious targets and secure communication with legitimate users.
& AO, SDR, and a low-complexity heuristic-based method.\\ \hline

\end{tabular}
\label{tableisac}
\end{table*}

\subsection{Takeaways}

In examining the extension of PA-ISAC systems, several important observations emerge regarding practical design constraints and system behavior. These points underscore the additional considerations that arise when enhancing PASS architectures with sensing functionality. Essentially, one can highlight the following observations:
\begin{itemize}
    \item The integration of target-sensing capabilities in PASS-ISAC systems often requires deploying additional transmitting and receiving waveguides extended from the same physical unit. This design choice enhances sensing performance but introduces extra hardware demands, thereby increasing the overall deployment cost. A careful assessment of this cost–performance trade-off is therefore necessary when considering sensing-augmented PASS architectures.
    \item The coexistence of adjacent transmitting and receiving waveguides may introduce a form of self-interference at the ISAC transceiver. Such interference can deteriorate both communication and sensing performance if not properly managed. Consequently, it must be explicitly modeled, mitigated, and incorporated into the optimization framework of PASS-ISAC to ensure reliable joint operation.
\end{itemize}

\section{Pinching Antenna Systems for Cutting-Edge Technologies}
\label{passcutedge}

The authors of \cite{passfl1} provided a visionary overview of the interplay between PASS and artificial intelligence (AI) schemes, such as federated learning (FL) or extended reality (XR). Notably, PASS can help in mitigating LoS blockage, improve the wireless channel for a better over-the-air computing, and enhance model misalignment in FL model updates. An initial investigation demonstrated that PA can enhance FL learning accuracy by providing a reliable channel to straggler FL clients. In \cite{passpls2}, the authors conducted a comprehensive analysis of a PA-aided FL system, considering both the synchronous FL (SFL) and asynchronous FL (AFL) update strategies. The adopted PASS is motivated by the need to enhance the channel quality of straggler nodes within an FL-based network. Analytical results demonstrated that an FL-aided PASS (FL-PASS) can achieve a shorter straggler offset distance in SFL compared to a conventional MIMO setup. Furthermore, results unveiled a notable gain of PA in terms of the average count of FL participants during a time period for AFL while getting rid of long-tail delays. The same authors proposed in \cite{passfl3} a robust FL-PASS scheme based on optimizing PA positions as well as client participation to minimize the expected time-to-accuracy metric. The position of the single radiating PA is then determined by partitioning the waveguide's length and proceeding with a single-dimensional root-finding process on the smoothed wall-clock objective function over a waveguide sub-interval. Observable gains were noted for the proposed PA placement and node participation scheme compared to baseline random node participation with PASS or with respect to a conventional MIMO scheme. In \cite{passfl4}, a straggler-resilient FL-PASS scheme was proposed and analyzed. By adopting a single-PA setup, the scheduled clients are grouped into three categories, namely: (i) pinching clients, uploading their updates by virtue of the PA, (ii) conventional clients transmitting their updates directly to a receive antenna at the server, and (iii) discarded clients. Clients belonging to each of the two categories are multiplexed via a NOMA scheme, while both categories of users are assigned different frequency bands to avoid interference. A fuzzy-logic-based scheme was adopted for clustering FL users per the above categories, considering per-user data contribution and channel characteristics. Then, for determined clusters of nodes, a DRL algorithm is deployed for identifying the optimal PA location, transmit power levels, and computational frequency, offering the minimal training latency. The proposed hybrid clustering and DRL-assisted PA positioning resulted in notable FL model accuracy gains against traditional antenna- and OMA-based baseline models.

\section{Open Problems and Directions Forward}
\label{passchallenges}

The previous sections provided a thorough system-wise review on the work performed on PASS over various system designs, where utilized optimization technique categories as well as their advantages and limitations were discussed. While several observations were emphasized on existing trade-offs and challenges in PASS deployment, there exists several challenges. This section provides an overview over some of these challenges, namely arising from the forecast deployment of PASS technology and its integration with existing wireless network standards.

\subsection{PASS and Other Channel Reconfiguring Techniques: Allies or Rivals?}

The fundamental objective of PASS lies in reshaping the wireless channel to the needs in communication, coverage, sensing, or secrecy. Along with PASS, RIS's use, prototype development, and field test was demonstrated in several research efforts over the past few years. Such findings demonstrated that RIS can provide observable gains in both indoor and outdoor wireless networks \cite{risexp,risexp2}. RIS beamsteering capability renders it efficient in creating virtual LoS whenever the direct link is obstructed. Intuitively, while, from one aspect, PASS and RIS can be viewed as two different solutions for the same propagation challenges, they can be also perceived as cooperating techniques which can jointly, in some cases, mitigate the LoS absence and increase the received power. For instance, the a joint cooperation between RIS and PASS is promising in scenarios where a LoS is probabilisitic or when the network is dynamic, where blocking objects change in location and dimensions. Thus, a hybrid RIS-PASS scheme can dynamically align the RIS phase shift with a given set of PAs not benefitting from a LoS link. Herein, PA position adjustment can be done to enhance the PA-RIS link, reducing the cascaded FSPL in RIS-aided systems.

The amalgamation of PASS with RIS has been recently analyzed in \cite{passris} for a multi-user network, which demonstrated a remarkable sum-rate gain compared to a baseline PASS. However, such an integration of both techniques requires (i) identifying underlying scenarios and use-cases in which the use of both techniques provides observable performance gains. In addition, since activating multiple reflective elements, in addition to the set of PAs, results in an additional energy consumption, it is crucial to consider an optimal operating strategy balancing the achievable performance gain and EE on hybrid PASS-RIS schemes.

\subsection{Extendability to Higher Frequency Bands}

The leaky waveguide theory formalizes the electromagnetic propagation behavior for PASS, demonstrating the potential radiation of the electromagnetic wave from a customized point of interest. It should be noted that although the majority of work on PASS consider a lossless waveguide, an actual waveguide features a distance-dependent signal power loss, which depends on the operating carrier frequency, waveguide material and structure \cite{passdirectional,passatten}. In lossy waveguides propagation model, the attenuation coefficient is proportional to the carrier frequency, which renders such an attenuation significant when implementing PASS on higher frequency bands, e.g., mmWave, sub-Terahertz (THz), or the THz band \cite{pozar,waveguideatten}. Such a challenging limitation calls for robust waveguide engineering that can cope with a larger in-waveguide loss for higher carrier frequencies.

From another front, the extension of PASS into the optical spectrum can attract a notable interest, due to \begin{enumerate}
    \item The current advance on optical wireless communication (OWC), being evaluated and demonstrated over a wide range of communication scenarios, and
    \item The established theory of leaky waveguides for the optical spectrum bands \cite{optwaveguide}.
\end{enumerate}   
To this end, performing a thorough analysis on the achievable gain of PASS at higher RF and optical bands is worth exploring as a future research direction.

\subsection{PA Positioning Mechanisms for Real-Time Application}

The underlying design of PASS can ensure a tunable channel response per the need for a given network setup. On the one hand, an optimization of the PA locations for each network configuration, e.g., channel realization, can ensure remarkable gains compared to traditional collocated antenna systems. On the other hand, an inherent challenges in PASS lies in the efficiently controlling the PA positioning for a dynamic network. In such a scenario, dynamic changes in the number of nodes, targets, or eavesdroppers, or even their position, require an real-time adaptation of PA positions, which may increase the mechanical cost and, consequently, the overall cost of deployment. Among the promising approaches to remedy to this issue, one can think of adopting a PASS optimized to maximize the erdogic performance rather than the instantaneous one, given a certain knowledge about the network dynamics. Another approach is the adoption of an on-off binary activation PASS, based on positioning a set of PAs in given positions and activating a subset of antennas per the instantaneous network and channel. 

While the vast majority of the reviewed schemes overlook real-time PA positioning issues in dynamic networks, it is understood that adaptive low-complexity PA positioning approaches are worth exploring in future works. Furthermore, it is crucial to investigate the EE of real-time PA positioning schemes, taking into account the mechanical/electrical cost from PA displacement.

\subsection{Channel State Information Acquisition}

The acquisition of the CSI of PASS manifests a particularity. In particular, PASS' channel estimation mechanism can differ than traditional MISO or MIMO schemes with collocated antennas, due to the potential presence of a higher number of antennas when adopting multiple waveguides with various PAs pinched on each. The unique channel matrix structure renders the estimation of the wireless channel matrix facing a rank-deficiency problem, irrespective of the pilot symbol length \cite{tutorial1}. Thus, some adaptive methods were developed for CSI estimation on PASS, such as sequential waveguide activation-based channel estimation. While such a strategy can combat the rank deficiency, it may results in a potential channel outdatedness in dynamic network. For instance, the estimated channel vector of the first waveguide can become outdated at the moment of transmission, taking place after estimating all the other waveguides' channel vectors, due to dynamics in the propagation environment. Therefore, developing robust channel estimation mechanisms and beamforming optimization schemes coping with the CSI aging are worth exploring as future directions.

\subsection{Integration with Wireless Network Infrastructure}

While current releases of wireless network standards, such as the third generation partnership program (3GPP), do not feature anything related to movable antennas or PAs, a potential integration of PAs with the current wireless standards requires a careful assessment of it real-time applicability for the various classes of service in the 5G and beyond. 

With the current momentum gained by PASS as a viable solution of the wireless channel-induced limitations in traditional MIMO schemes, one can forecast a cooperative aspect between PASS and conventional collocated antenna-based schemes. For instance, both collocated antenna arrays and waveguides can be linked to the same BS, where a hybrid operation between both setups (e.g., collocated and PASS) can be performed per the observed channel conditions with respect to the served nodes. Nonetheless, such a hybrid architecture that garners the benefits of both techniques, can face some implementation challenges. For instance, as detailed in the previous subsection, due to the difference in (i) the number of antennas and (ii) the CSI estimation procedure between PASS and conventional systems, a different pilot signaling structure can be manifested for both. Thus, this calls for the need to develop a unified pilot signaling and channel estimation technique, able to retrieve the wireless channel matrices for both the collocated MIMO sub-system and the PASS one.

From an evaluation perspective, an interesting direction to consider is the analysis of the potential of hybrid communication and/or sensing networks composed of PASS and conventional multi-antenna arrays. For instance, the authors in \cite{passhybrid} analyzed recently a cooperative scheme involving both aforementioned techniques, where three cooperation strategies were proposed and analyzed. Such hybrid schemes can be extended to a dynamic network with mobile users and dynamic blockers or variable size and location, which requires a robust cooperation between both subsystems.

\section{Conclusions}
\label{passconclusion}

In this survey, a system-wise review of the work conducted on PASS was performed. In particular, an overview of the PASS technique was presented and detailed, including the propagation model, signal and channel representation, and the various PASS deployment architectures at the hardware level. Then, the analyzed work from the body of literature on PASS designs was viewed from the perspective of different attainable objectives, namely on communication, sensing, and secure transmission. Such a system-wise review provided considerable insights into the achievable gains of the various PASS architectures, including single- and multi-waveguide ones, on the aforementioned objectives. Additionally, several observations were provided on the limitations of the various PASS schemes reviewed. 
Lastly, a thorough discussion on the foreseen PASS implementation challenges was provided, along with pointing out interesting future directions to investigate in the topic.

\bibliographystyle{IEEEtran}
\bibliography{refs}

\end{document}